\documentclass{article}

\usepackage[utf8]{inputenc}

\usepackage[T1]{fontenc}
\usepackage{parskip}
\usepackage{ragged2e}
\usepackage{enumitem}
\usepackage{float}
\usepackage{lscape}
\usepackage{rotating}

\usepackage{fancyhdr}
\usepackage{etoolbox}
\usepackage{geometry}
 
\usepackage{amsmath}
\usepackage{amsfonts}
\usepackage{amssymb}
\usepackage{mathtools}
\usepackage{amsthm}
\usepackage{bbm}
\usepackage{nccmath}
\usepackage{scalerel}
\usepackage{actuarialsymbol}
\usepackage{mleftright}

\usepackage{multirow}
\usepackage{multicol}
\usepackage{tabularx}

\usepackage{graphicx}
\usepackage[table]{xcolor}
\usepackage{array}
\usepackage{subcaption}

\usepackage{xurl}
\usepackage{thmtools, thm-restate}

\usepackage[sectionbib, round]{natbib}
\usepackage[colorlinks, linkcolor = black,  linkcolor = blue,
urlcolor  = blue,
citecolor = blue,
anchorcolor = blue]{hyperref}
\usepackage{etoolbox}
  
\makeatletter
\newcounter{author}
\renewcommand*\author[1]{%
  \stepcounter{author}%
  \ifnum\c@author=1
    \gdef\@author{#1}%
  \else
    \xdef\@author{\unexpanded\expandafter{\@author\and#1}}%
  \fi
  \csgdef{author@\the\c@author}{#1}}
\newcommand*\email[1]{%
  \csgdef{email@\the\c@author}{#1}}
\newcommand*\orcid[1]{%
  \csgdef{orcid@\the\c@author}{#1}}
\newcommand*\address[1]{%
  \csgdef{address@\the\c@author}{#1}}
\AtEndDocument{%
  \xdef\author@count{\the\c@author}%
  \c@author=1
  \print@authors}
\newcommand*\print@authors{%
  \ifnum\c@author>\author@count
  \else
    \print@author{\the\c@author}%
    \advance\c@author by 1
    \expandafter\print@authors
  \fi}
\newcommand*\print@author[1]{%
  \par\medskip
  \begin{tabular}{@{}l@{}}%
    \textsc{\csuse{author@#1}}\\
    \csuse{address@#1}\\
    \textit{E-Mail}:
    \href{mailto:\csuse{email@#1}}{\csuse{email@#1}}\\
    \textit{ORCiD}:
    \href{\csuse{orcid@#1}}{\csuse{orcid@#1}}
  \end{tabular}}
\patchcmd{\NAT@test}{\else \NAT@nm}{\else \NAT@hyper@{\NAT@nm}}{}{}
\makeatother

\numberwithin{equation}{section}

\newtagform{normalsize}[\normalsize]{\normalsize(}{\normalsize)}

\usetagform{normalsize}

\theoremstyle{definition}
\newtheorem{definition}{Definition}
\numberwithin{definition}{section}
\theoremstyle{plain}
\newtheorem{proposition}{Proposition}
\numberwithin{proposition}{section}
\theoremstyle{plain}
\newtheorem{corollary}{Corollary}
\numberwithin{corollary}{section}
\theoremstyle{remark}
\newtheorem{remark}{Remark}
\numberwithin{remark}{section}
\mathtoolsset{showonlyrefs}

\providecommand{\keywords}[1]
{
  \small	
  \textbf{\textit{Keywords---}} #1
}

\title{Subsidising Inclusive Insurance to Reduce Poverty}

\author{Jos\'e Miguel Flores-Contró}
\address{\textit{Department of Actuarial Science} \\ \textit{Faculty of Business and Economics} \\
\textit{University of Lausanne} \\ \textit{Lausanne, Switzerland}}
\email{josemiguel.florescontro@unil.ch}
\orcid{https://orcid.org/0000-0001-9332-8811}
\author{Kira Henshaw}
\address{\textit{Institute for Financial and Actuarial Mathematics} \\ \textit{Department of Mathematical Sciences} \\ \textit{University of Liverpool} \\ \textit{Liverpool, United Kingdom}}
\email{kirahenshaw@gmail.com}
\orcid{https://orcid.org/0000-0002-9018-7993}
\author{Sooie-Hoe Loke}
\address{\textit{Department of Mathematics} \\ \textit{Central Washington University} \\ \textit{Ellensburg, Washington} \\ \textit{United States of America}}
\email{sooiehoe.loke@cwu.edu}
\orcid{N/A}
\author{S\'everine Arnold}
\address{\textit{Department of Actuarial Science} \\ \textit{Faculty of Business and Economics} \\
\textit{University of Lausanne} \\ \textit{Lausanne, Switzerland}}
\email{severine.arnold@unil.ch}
\orcid{https://orcid.org/0000-0001-9851-6328}
\author{Corina Constantinescu}
\address{\textit{Institute for Financial and Actuarial Mathematics} \\ \textit{Department of Mathematical Sciences} \\ \textit{University of Liverpool} \\ \textit{Liverpool, United Kingdom}}
\email{c.constantinescu@liverpool.ac.uk}
\orcid{https://orcid.org/0000-0002-5219-3022}

\begin{document}

\date{\vspace{-2ex}}

\maketitle

\begin{abstract}
In this article, we assess the benefits of coordination and partnerships between governments and private insurers, and provide further evidence for microinsurance products as powerful and cost-effective tools for achieving poverty reduction. To explore these ideas, we model the capital of a household from a ruin-theoretic perspective to measure the impact of microinsurance on poverty dynamics and the governmental cost of social protection. We analyse the model under four frameworks: uninsured, insured (without subsidies), insured with subsidised constant premiums and insured with subsidised flexible premiums. Although insurance alone (without subsidies) may not be sufficient to reduce the likelihood of falling into the area of poverty for specific groups of households, since premium payments constrain their capital growth, our analysis suggests that subsidised schemes can provide maximum social benefits while reducing governmental costs.

\vspace{0.5cm}


\end{abstract}

\section{Introduction} \label{Introduction-Section1}

In recent years, governments in developing countries have been increasingly involved in the provision of insurance programmes. In countries such as China and India for instance, the agricultural insurance sector has grown significantly thanks to the support (and premium subsidies) provided by central and provincial governments \citep{Article:Kramer2022}. While doubts about the role of insurers in alleviating poverty exist among practitioners, adequate coordination between governments, private insurance companies and other stakeholders (e.g. NGOs, international financial institutions and other donors) has been shown to enhance the development of sustainable, affordable and cost-effective insurance products \citep{Article:Linnerooth-Bayer2007, Article:Auzzir2014}. 

The most common form of government support for insurance are premium subsidies. The central and provincial governments provide Chinese farmers with subsidies exceeding $50\%$ of the premium amount, of which farmers pay only about $20\%$ \citep{Article:Wang2011, Article:Ye2020}. Similarly, the Pradhan Mantri Fasal Bima Yojana (PMFBY), a government-sponsored multi-stakeholder crop insurance scheme in India, charges farmers a maximum premium ranging from $2\%$ to $5\%$ of the sum insured (or the actuarial rate, whichever is lower), with the remaining part of the premium paid on a 50/50 basis by the central and state governments \citep{Article:Kaur2021}. Insurance premium subsidies must be designed with a clearly stated purpose. They should target those in need and address market deficiencies or consumer equity concerns (see \cite{Book:Hill2014}, which is a technical report from the United Nations\rq \ International Labour Organization (ILO)). Experience shows that, when designed properly, subsidised insurance schemes represent a powerful and cost-effective way to achieve public policy objectives, while poorly designed insurance premium subsidies can be inefficient and lead to significant economic costs \citep{Article:Hazell2020}. 

Adopting the novel ruin-theoretic approach presented by \cite{Article:Kovacevic2011}, this article studies the impact of insurance (both with and without subsidies) on poverty dynamics and the governmental cost of social protection. Through this analysis, we seek to determine the benefits derived from coordination and partnerships between governments and private insurers, and to highlight the cost-effectiveness of government support for insurance. Previous studies have approached the same problem from a dynamic stochastic programming perspective. \cite{Inbook:Barrett2016}, \cite{Article:Carter2018} and \cite{Article:Janzen2020} propose dynamic models of household consumption, investment and risk management, considering a social insurance-type mechanism which first prioritises lending aid to the vulnerable non-poor, contingent on their experience of negative shocks, then to those already below the poverty line. Introduction of an index-based insurance market is found to outperform the asset-based vulnerability-targeted protection in poverty reduction, economic growth and the cost of social protection. Although implementation of a vulnerability-targeted strategy induces a short-term increase in poverty, rates are lower than those associated with both in-kind and cash transfers in the medium- and long-term.

\cite{Article:Carter2018} and \cite{Article:Janzen2020} compare the impact of insurance when all costs are paid by the policyholder and when targeted-subsidies are provided to the vulnerable and already poor. In the latter study, those in the neighbourhood of the poverty line do not optimally purchase insurance (without subsidies), instead suppressing their consumption and mitigating the probability of falling into poverty. Observing a greater reduction in poverty in comparison to pure cash transfers, \cite{Article:Jensen2017} provide empirical evidence for the benefits of insurance-based social protection through analysis of safety net and drought-based livestock insurance programmes in northern Kenya. \cite{Article:Chantarat2017} consider the welfare impacts of the same index-based insurance programme, using herd size dynamics to address the vulnerability to poverty associated with covariate livestock mortality such that critical herd size mimics the poverty line. Here, targeted premium subsides are optimised across various herd size groups such that given measures of poverty reduction are maximised. Increases and decreases in household wealth and poverty, respectively, were greater under the optimal strategy than under alternative needs-based subsidisation mechanisms and with no insurance. In the presence of needs-based subsidisation which provides free protection to the most poor, the number of poor continued to increase, thus highlighting the importance of social protection strategies that target those still above but close to the poverty line in addition to the already poor.

The insurance strategies considered in these studies are inclusive insurance mechanisms specifically designed to cater for the most vulnerable. Inclusive insurance, commonly referred to as microinsurance, relates to the provision of insurance services to low-income populations with limited, or no access to mainstream insurance or alternative effective risk management strategies. Targeting low-income individuals living close to or below the poverty line, microinsurance aims to close the protection gap that exists between uninsured and insured losses to life, property and health by providing protection to the poor. However, barriers to microinsurance penetration exist due to constraints on product affordability resulting from fundamental features of the microinsurance environment. These distinct features include the nature of low-income risks, limited consumer financial literacy and experience, product accessibility and data availability. While novel solutions for the supply and distribution of products in this environment exist (e.g. mobile-based business models \citep{Article:Kousky2021}), it is important to consider the viability of microinsurance uptake for all sectors of the target population, particularly for the most vulnerable.

Premium payments can in fact heighten the risk of falling into poverty for the proportion of the population living just above the poverty line, inducing a balance between protection and loss as a result of insurance coverage which is dependent on the entity's level of capital (see, for example, \cite{Article:Kovacevic2011} and \cite{Article:Liao2020}, where the latter use a multiple-equilibrium framework to analyse the impact of subsidised and unsubsidised agricultural insurance on poverty rates in rural China). This insufficiency of microinsurance alone as a means for poverty reduction for the most exposed necessitates an alternative solution. For this purpose, as in the aforementioned studies, we consider microinsurance schemes which are supported by social protection strategies, and more specifically, their potential in minimising both the probability of a household falling below the poverty line and the governmental cost of social protection. For thorough discussions of microinsurance, the challenges associated with adapting commercial insurance to serve the poor and the insurability of risks in the market, the interested reader may refer to \cite{Article:Dror2019}, \cite{Article:Churchill2007} and \cite{Article:Biener2012}.

Besides reducing the impact on household capital growth, the use of subsidies to lower consumer premium payments has the potential to increase microinsurance uptake, with wealth and product price positively and negatively influencing microinsurance demand, respectively, see \cite{Article:Eling2014} and \cite{Article:Platteau2017}. However, this relationship is not transparent (see, for instance, \cite{Article:Cole2013}, where the authors find that, in the city of Ahmedabad in India, more than half of households in their sample do not purchase rainfall insurance even when premiums are set significantly below actuarially fair values), with additional factors, including financial education levels, insurer trust and logistical problems in the purchase and renewal of coverage, having significant influence on a household’s decision to insure.  A comprehensive approach should therefore be adopted by insurance providers such that low-cost subsidisation schemes are complemented by innovative activities improving understanding of and access to insurance products. Focusing specifically on agricultural insurance, \cite{Article:Hazell2020} present government and donor incentives for subsidisation. As an example, temporary subsidies can enable low-income farmers to bear the risk of adopting innovative technologies which may bring them out of poverty. However, in addition to improving the economic circumstances of the insured, through the provision of insurance experience this strategy mitigates the uncertainties surrounding insurance common among consumers in the microinsurance environment, while improving the quality of consumer data. The study additionally highlights how subsidisation schemes help to scale up insurance products.

Although important for poverty alleviation, the behaviour of a household below the poverty line is not considered in this study. Households that live or fall below the poverty line are said to be in a poverty trap, where a poverty trap is a state of poverty from which it is difficult to escape without external help. Poverty trapping is a well-studied topic in development economics (the interested reader may refer to \cite{Inbook:Aghion2005}, \cite{Book:Bowles2006}, \cite{Article:Kraay2014}, \cite{Article:Barrett2016} and references therein for further discussion; see \cite{Inbook:Matsuyama2008} for a detailed description of the mechanics of poverty traps), however, for the purpose of this study, we use the term \lq \lq trapping\rq \rq \ only to describe the event that a household falls into poverty, focusing our interest on low-income behaviours above this critical line.

Our study complements the aforementioned studies that analyse the impact of inclusive insurance from both an empirical and dynamic stochastic programming perspective. We introduce a more formal and rigorous mathematical framework that analytically demonstrates the benefits of partnerships between governments and private insurers. For that purpose, we adapt the piecewise-deterministic Markov process proposed by  \cite{Article:Kovacevic2011} such that households are subject to shocks of random size and we consider a non-discretised capital process. In line with the poverty trap ideology, we assume the area of poverty to be an absorbing state and consider only the state of events above the poverty threshold. Obtaining explicit solutions for the trapping probability and the governments\rq \ cost of social protection using classical risk theory techniques (where this is considered the ideal scenario \citep{Book:Asmussen2010}), we compare the influence of three structures of microinsurance on these quantities. Specifically, we consider a microinsurance scheme with (i) unsubsidised premiums, (ii) subsidised constant premiums and (iii) subsidised flexible premiums. Unlike previous studies, where the cost of social protection is defined as the present value of government subsidies plus the transfers needed to close the poverty gap for all poor households (see, for example, \cite{Inbook:Barrett2016} and \cite{Article:Janzen2020}), the ruin-theoretic perspective adopted in the proposed model allows us to include a supplemental fixed cost that ensures, with a certain level of confidence, that households will not return to poverty, should they fall underneath the threshold. In this way, the likelihood that the government will re-incur these costs for the same household is reduced.

The adopted capital models are special cases of well-studied risk theory models. Therefore, a standard modelling approach with application to poverty trapping is considered. Typically assumed to represent the surplus process of an insurer, our alternative application enforces two key adaptations. First, unlike the barrier at zero considered in the classical setting, where an insurer is deemed to be ruined if their surplus falls below zero, a non-zero critical barrier reflecting the poverty line is assumed. Second, in the classical case, insurers raise capital and rely on access to reinsurance to avoid falling below the critical level. Thus, two mechanisms for escaping ruin exist. On the other hand, the level of capital growth attained by a household is the only mechanism protecting them from ruin in the absence of insurance. In addition, while in the classical setting the initial surplus is typically considered to be large enough to keep the insurer away from ruin, in this study we focus on analysing households with initial capital just above the poverty line.

Under the first premium framework we assume premium payments are made by households. We demonstrate that these payments can constrain households\rq \ capital growth and thus increase their trapping probability compared to that of uninsured households, as previous studies have shown. Conversely, under such a scheme, the cost of social protection remains lower than the corresponding uninsured cost. With the need for an alternative solution to address the observed negative impact on poverty dynamics, under the second premium framework we assume governments provide insurance premium subsidies to all households. Reducing premium payments by means of subsidies has a positive impact on household capital growth and their trapping probabilities. Furthermore, the results obtained allow us to estimate optimal subsidies for households with varying degrees of capital such that they preserve a trapping probability equal to that of when uninsured. The proposed subsidy optimisation aligns with the idea of \ \lq \lq smart\rq \rq \ subsidies, which provide maximum social benefits while minimising distortions in the insurance market and the mis-targeting of clients \citep{Book:Hill2014}. Here, the optimal subsidy seeks to reduce the likelihood of a household falling into the area of poverty (clear objective), has a mathematical foundation (transparent), intends to help those in need of assistance (targeted), can be assessed over time (monitoring and evaluation), can be strategically planned (exit strategy/long-term financing) and is capable of being costed (costs contained). Our analysis shows that, under this subsidised microinsurance scheme, while government support is not essential for privileged households, vulnerable households with capital levels close to the poverty line require assistance. Moreover, the cost of social protection is lower for the most vulnerable than in the corresponding uninsured framework, but is higher for the most privileged.

Mimicking the well-known risk theory dividend barrier strategy, the third framework considers a novel scheme where households pay premiums only when their capital is above some pre-defined capital barrier, with the premium otherwise paid by the government. Granting flexibility on premium payments allows households to attain lower trapping probabilities, since they are assisted by the government when their capital lies close to the poverty trap. Continuing with the idea of \lq \lq smart\rq \rq \ subsidies, we optimise the capital barrier level at which governments should begin providing support. As could be expected, those closest to the poverty line require immediate aid, with optimal barriers lying above their initial capital, whereas those further away from the poverty trap possess the ability to pay premiums themselves once enrolled in the scheme, yielding to optimal barriers lying below their initial capital levels. Under this framework, the cost of social protection remains lower than the corresponding uninsured cost.

Premium subsidies are not phased out over time in the inclusive insurance schemes considered here. Nevertheless, it is often necessary to assess the financial dependence of subsidised inclusive insurance schemes on external support. That is, situations in which governments decide to reduce or end the provision of subsidies, that may lead to the need to raise premiums beyond the reach of their customers, should be taken into account when evaluating the viability of a subsidised scheme, as they expose concerns regarding the scheme\rq s sustainability. 

Informal risk-sharing networks are highly prevalent in low-income economies (see, for example, \cite{Article:Townsend1994}, \cite{Book:Bardhan1999}, \cite{Article:DeWeerdt2006}). In particular, community-based networks in Ghana and South Africa gather funds and other contributions to meet funeral expenses \citep{Article:Ramsay2013}. These networks help to mitigate the risk of idiosyncratic losses. Heavily subsidising an insurance product would, however, lessen the need for such networks, as policyholders would be protected from the occurrence of adverse events. Hence, the strength of the networks could suffer as a result, leaving households exposed if the subsidy is eventually removed. Although our definition of \lq \lq insurance\rq \rq \ covers all forms of losses, in reality, an insurance policy typically covers a single peril. It is therefore likely that, even with subsidised insurance, low-income households will still participate in risk-sharing mechanisms to mitigate the losses that are not covered by their insurance policies. Thus, considering the situation in which subsidies cease, the informal networks could be rebuilt on the foundations that remain. Furthermore, if phasing out subsidies, governments could undertake activities to encourage the continuation of informal risk-sharing to prevent the crowding out of informal risk-transfer mechanisms. Evidence does, however, suggest that insurance uptake drops when a subsidy is removed \citep{Article:Platteau2017}, implying that households revert back to informal risk-sharing mechanisms rather than purchasing unsubsidised coverage. Formal insurance and informal risk-sharing networks have, in fact, been found to be complementary (see, for example, \cite{Article:Will2021}).

Government subsidies may induce an increase in moral hazard, since the reduction in premiums diminishes policyholders\rq \ sensitivity to the consequences of a loss. However, as described by \cite{Article:Biener2012} and \cite{Book:Hill2014} among others, reducing information asymmetries through government investment in data collection can help to alleviate the increased risk by enabling better understanding of a policyholder’s true risk exposure. This issue is of particular concern in the health, agricultural and catastrophe insurance markets. Distribution through local enterprises \citep{Article:Dercon2006}, group-based products \citep{Article:Biener2012} and financial education \citep{Article:Biener2014} have also been found to lessen information asymmetries associated with moral hazard.

The remainder of the paper is structured as follows. In Section \ref{TheCapitalModel-Section2}, we introduce the household capital model and its associated infinitesimal generator. The (trapping) time at which a household falls into the area of poverty is defined in Section \ref{TheTrappingTime-Section3}, and subsequently the explicit trapping probability and the expected trapping time are derived for the basic uninsured model. Links between classical ruin models and the trapping model of this paper are stated in Sections \ref{TheCapitalModel-Section2} and \ref{TheTrappingTime-Section3}. Microinsurance is introduced in Section \ref{IntroducingMicroinsurance-Section4}, where we assume a proportion of household losses are covered by a microinsurance policy. The capital model is redefined and the trapping probability is derived. Sections \ref{MicroinsurancewithSubsidisedConstantPremiums-Section5} and \ref{MicroinsurancewithSubsidisedFlexiblePremiums-Section6} consider the case where households are proportionally insured through a government subsidised microinsurance scheme, with the impact of subsidised flexible premiums discussed in Section \ref{MicroinsurancewithSubsidisedFlexiblePremiums-Section6}. Optimisation of the subsidy and capital barrier levels are presented in Sections \ref{MicroinsurancewithSubsidisedConstantPremiums-Section5} and \ref{MicroinsurancewithSubsidisedFlexiblePremiums-Section6}, alongside the associated governmental cost of social protection. Concluding remarks are provided in Section \ref{Conclusion-Section6}.

\section{The Capital Model} \label{TheCapitalModel-Section2}

The fundamental dynamics of the model follow those of \cite{Article:Kovacevic2011}, where the growth in accumulated capital $(X_t)$ of an individual household is given by 

\vspace{0.3cm}

\begin{align}
    \frac{dX_{t}}{dt}=r \cdot\left[X_{t}-x^{*}\right]^{+}, \label{TheCapitalModel-Section2-Equation1}
\end{align}

\vspace{0.3cm}

where $[x]^{+}=\max(x,0)$. The capital growth rate $r = (1-a) \cdot b \cdot c > 0$ incorporates household rates of consumption ($0<a<1$), income generation ($0<b$) and investment or savings ($0<c<1$), while $x^* > 0$ represents the threshold below which a household lives in poverty. Reflecting the ability of a household to produce, accumulated capital $(X_t)$ is composed of land, property, physical and human capital, with health a form of capital in extreme cases where sufficient health services and food accessibility are not guaranteed \citep{Article:Dasgupta1997}. The notion of a household in this model setting may be extended for consideration of poverty trapping within economic units such as community groups, villages and tribes, in addition to the traditional household structure.

The dynamical process in \eqref{TheCapitalModel-Section2-Equation1} is constructed such that consumption is assumed to be an increasing function of wealth (for full details of the model construction see \cite{Article:Kovacevic2011}). The  poverty threshold $x^{*}$ represents the amount of capital required to forever attain a critical level of income, below which a household would not be able to sustain their basic needs, facing elementary problems relating to health and food security. Throughout the paper, we will refer to this threshold as the critical capital or the poverty line. Since \eqref{TheCapitalModel-Section2-Equation1} is positive for all levels of capital greater than the critical capital, points less than or equal to $x^{*}$ are stationary (capital remains constant if the critical level is not met). In this basic model, stationary points below the critical capital are not attractors of the system if the initial capital exceeds $x^{*}$, in which case the capital process $(X_{t})$ grows exponentially with rate $r$.

Using capital as an indicator of financial stability over other commonly used measures such as income enables a more effective analysis of a household\rq s wealth and well-being. Households with relatively high income, considerable debt and few assets would be vulnerable if any loss of income was to occur, while low-income households could live comfortably on assets acquired during more prosperous years for a long-period of time \citep{Book:Gartner2004}.

In line with \cite{Article:Kovacevic2011}, we expand the dynamics of \eqref{TheCapitalModel-Section2-Equation1} under the assumption that households are susceptible to the occurrence of capital losses, including severe illness, the death of a household member or breadwinner and catastrophic events such as floods and earthquakes. We assume occurrence of these events follows a Poisson process with intensity $\lambda$, where the capital process follows the dynamics of \eqref{TheCapitalModel-Section2-Equation1} in between events. On the occurrence of a loss, the household's capital at the event time reduces by a random amount $Z_{i}$. The sequence $(Z_{i})$ is independent of the Poisson process and i.i.d. with common distribution function $G_{Z}$. In contrast to \cite{Article:Kovacevic2011}, we assume reduction by a given amount rather than a random proportion of the capital itself. This adaptation enables analysis of a tractable mathematical model that provides, for instance, the possibility of finding an analytical solution for the infinite-time trapping probability (defined in Section \ref{TheTrappingTime-Section3}). This differs from previous work in which numerical methods, considered by \cite{Book:Asmussen2010} as the second best alternative to calculating trapping probabilities when closed-form expressions are not available, are employed to estimate such a quantity (see, for example, \cite{Article:Kovacevic2011} and \cite{Article:Azais2015}). The core objective of studying the probability of a household falling into the area of poverty remains. 

A household reaches the area of poverty if it suffers a loss large enough that the remaining capital is attracted into the poverty trap. Since a household's capital does not grow below the critical capital $x^{*}$, households that fall into the area of poverty will never escape without external help. Once below the critical capital, households are exposed to the risk of falling deeper into poverty, with the dynamics of the model allowing for the possibility of negative capital. A reduction in a household's capital below zero could represent a scenario where total debt exceeds total assets, resulting in negative capital net worth. The experience of a household below the critical capital is, however, out of the scope of this paper.

We will now formally define the stochastic capital process, where the process for the inter-event household capital \eqref{TheCapitalModel-Section2-Equation2} is derived through solution of the first order Ordinary Differential Equation (ODE) \eqref{TheCapitalModel-Section2-Equation1}. This model is an adaptation of the model proposed by \cite{Article:Kovacevic2011}.

\vspace{0.3cm}

\begin{definition} \label{TheCapitalModel-Section2-Definition1}
Let $T_{i}$ be the $i^{th}$ event time of a Poisson process $\left(N_{t}\right)$ with parameter $\lambda$, where $T_{0}=0 .$ Let $Z_{i} \ge 0 $ be a sequence of i.i.d. random variables with distribution function $G_{Z}$, independent of the process $\left(N_{t}\right)$. For $T_{i-1} \leq t<T_{i}$, the stochastic growth process of the accumulated capital $X_{t}$ is defined as

\vspace{0.3cm}

\begin{align}
    X_{t}=\begin{cases} \left(X_{T_{i-1}}-x^{*}\right) e^{r \left(t-T_{i-1}\right)}+x^{*} & \textit { if } X_{T_{i-1}}>x^{*}, \\ X_{T_{i-1}} & \textit{ otherwise.}
    \end{cases}
    \label{TheCapitalModel-Section2-Equation2}
\end{align}

\vspace{0.3cm}

At the jump times $t = T_{i}$, the process is given by

\vspace{0.3cm}

\begin{align}
    X_{T_{i}}=\begin{cases} \left(X_{T_{i-1}}-x^{*}\right) e^{r \left(T_{i}-T_{i-1}\right)}+x^{*} - Z_{i} & \textit { if } X_{T_{i-1}}>x^{*}, \\ X_{T_{i-1}} - Z_{i} & \textit{ otherwise.}
    \end{cases}
    \label{TheCapitalModel-Section2-Equation3}
\end{align}
\end{definition}

\vspace{0.3cm}

The infinitesimal generator of the stochastic process $(X_t)_{t\geq 0}$, which is a piecewise-determinsitic Markov process \citep{Article:Davis1984}, is given by

\vspace{0.3cm}

\begin{align}
    (\mathcal{A} f)(x)=r(x-x^{*}) f^{\prime}(x) +\lambda \int_{0}^{\infty} \left[f(x - z) - f(x)\right] \mathrm{d} G_{Z}(z) ,  \qquad x \ge x^{*}.
    \label{TheCapitalModel-Section2-Equation4}
\end{align}

\vspace{0.3cm}

The capital model as defined in \eqref{TheCapitalModel-Section2-Equation2} and \eqref{TheCapitalModel-Section2-Equation3} is in fact a topic well-studied in ruin theory since the 1940s. As such, well-established techniques can be easily applied to the poverty trapping context of this paper. In ruin theory, modelling is undertaken from the point of view of an insurance company. Consider the  insurer\rq s surplus process $(U_t)_{t\geq 0}$ given by 

\vspace{0.3cm}

\begin{align}
  U_t =u+pt+ \nu \int_0^t U_s \, ds-\sum_{i=1}^{N_t} Z_i,
   \label{TheCapitalModel-Section2-Equation5}
\end{align} 

\vspace{0.3cm}

where $u$ is the insurer\rq s initial capital, $p$ is the constant premium rate, $\nu$ is the risk-free interest rate, $N_t$ is a Poisson process with parameter $\lambda$ which counts the number of claims in the time interval $[0,t]$ and $(Z_i)_{i=1}^\infty$ is a sequence of i.i.d. claim sizes with distribution function $G_{Z}$. This model is also called the insurance risk model with deterministic investment, first proposed by \cite{Article:Segerdahl1942}  and subsequently studied by \cite{Article:Harrison1977} and \cite{Article:Sundt1995}. For a detailed literature review on this model prior to the turn of the century, readers may consult \cite{Article:Paulsen1998}. 

Observe that when $p=0$,  the  insurance model  \eqref{TheCapitalModel-Section2-Equation5} for positive surplus is equivalent to  the capital model  \eqref{TheCapitalModel-Section2-Equation2} and \eqref{TheCapitalModel-Section2-Equation3} above the poverty line $x^*=0$. Subsequently, the capital growth rate $r$ in our model corresponds to the risk-free investment rate $\nu$ of the insurer\rq s surplus model. More connections between these two models will be made in the next section following introduction of the trapping time. 

\section{The Trapping Time} \label{TheTrappingTime-Section3}
Let 

\vspace{0.3cm}

\begin{align}
    \tau_{x}:=\inf \left\{t \geq 0: X_{t}<x^{*} \mid X_{0}=x\right\}
    \label{TheTrappingTime-Section3-Equation1}
\end{align}

\vspace{0.3cm}

denote the time at which a household with initial capital $x \ge x^{*}$ falls into the area of poverty (the trapping time) and let $\psi(x) = \mathbb{P}(\tau_{x} < \infty)$ be the infinite-time trapping probability. To study the distribution of the trapping time we apply the expected discounted penalty function at ruin, a concept commonly used in actuarial science \citep{Article:Gerber1998}, such that with a force of interest $\delta \ge 0$ and initial capital $x \ge x^{*}$, we consider

\vspace{0.3cm}

\begin{align}
    m_{\delta}(x)= \mathbb{E}\left[w(X_{\tau^{-}_{x}}- x^{*},\mid X_{\tau_{x}}-x^{*}\mid)e^{-\delta \tau_{x}} \mathbbm{1}_{\{\tau_{x} < \infty\}}\right],
    \label{TheTrappingTime-Section3-Equation2}
\end{align}

\vspace{0.3cm}

where $\mathbbm{1}_{\{A\}}$ is the indicator function of a set $A$ and $w(x_{1}, x_{2})$ for $0 \leq x_{1},  x_{2} < \infty $, is a non-negative penalty function of $x_1$, the capital surplus prior to the trapping time, and $x_2$, the capital deficit at the trapping time. For more details on the so-called Gerber-Shiu risk theory, interested readers may wish to consult \cite{Book:Kyprianou2013}.

The probabilistic properties of the trapping time are contained in its distribution function. However, it is sometimes much easier to work with a transformation rather than with the distribution function of a random variable itself. Here, we focus on the Laplace transform, which is particularly useful for nonnegative, absolutely continuous random variables such as the trapping time and is a powerful tool in probability theory. Moreover, the Laplace transform characterises the probability distribution uniquely. For a continuous random variable $X$, with probability density function $f_{X}$, the Laplace transform of $f_{X}$ is given by the expected value $\mathcal{L}\{f_{X}\}\left(s\right)=\mathbb{E}\left[e^{-sX}\right]$. Note that, specifying the penalty function such that $w(x_{1}, x_{2})=1$ in  \eqref{TheTrappingTime-Section3-Equation2}, $m_{\delta}(x)$ becomes the Laplace transform of the trapping time, also interpreted as the expected present value of a unit payment due at the trapping time.

For simplicity, throughout the rest of the paper we will assume that capital losses are exponentially distributed $(Z_{i} \sim Exp(\alpha))$.

\vspace{0.3cm}

\begin{proposition}\label{TheTrappingTime-Section3-Proposition1}

Consider a household capital process (as proposed in Definition \ref{TheCapitalModel-Section2-Definition1}) with initial capital $x\ge x^{*}$, capital growth rate $r$, intensity $\lambda > 0$ and exponentially distributed capital losses with parameter $\alpha > 0$. The Laplace transform of the trapping time is given by

\vspace{0.3cm}

\begin{align}
    m_{\delta}(x)=\frac{\lambda}{(\lambda + \delta) U\left(1-\frac{\lambda}{r},1-\frac{\lambda+\delta}{r};0\right)}e^{y(x)}U\left(1-\frac{\lambda}{r},1-\frac{\lambda+\delta}{r};-y(x)\right),
    \label{TheTrappingTime-Section3-Equation3}
 \end{align}

\vspace{0.3cm}

where $\delta \ge 0$ is the force of interest for valuation, $y(x)=-\alpha (x-x^{*})$ and $U(\cdot)$ is Tricomi\rq s Confluent Hypergeometric Function as defined in \eqref{Appendix A: Mathematical Proofs-Equation8}. 

\end{proposition}

\vspace{0.3cm}

See Appendix \ref{ProofofProposition3.1} for proof of Proposition \ref{TheTrappingTime-Section3-Proposition1}.

\vspace{0.3cm}

\begin{figure}[H]
	\begin{subfigure}[b]{0.5\linewidth}
  		\includegraphics[width=8cm, height=8cm]{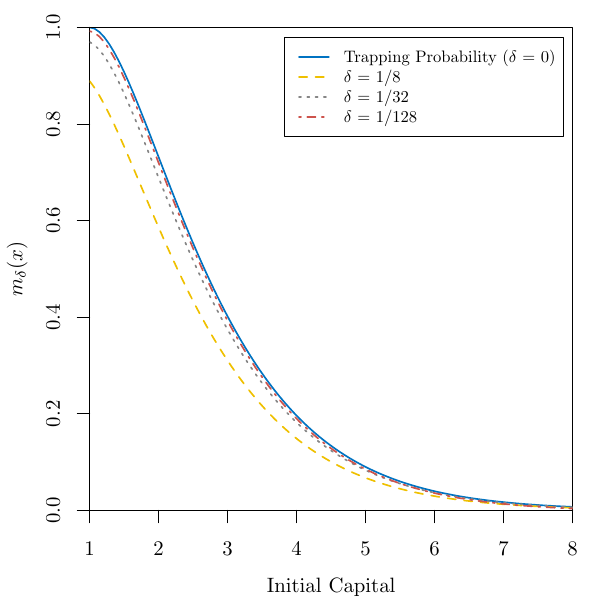}
		\caption{}
  		\label{TheTrappingTime-Section3-Figure1-a}
	\end{subfigure}
	\begin{subfigure}[b]{0.5\linewidth}
  		\includegraphics[width=8cm, height=8cm]{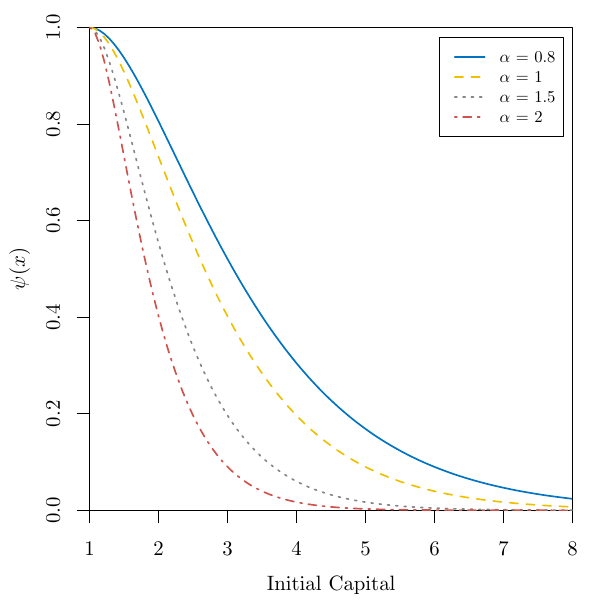}
		\caption{}
  		\label{TheTrappingTime-Section3-Figure1-b}
	\end{subfigure}
	\caption{(a) Laplace transform $m_{\delta}(x)$ of the trapping time when $Z_{i} \sim Exp(1)$, $a = 0.1$, $b = 1.4$, $c = 0.4$, $\lambda = 1$, $x^{*} = 1$ for $\delta = 0, \frac{1}{8}, \frac{1}{32}, \frac{1}{128}$ (b) Trapping probability $\psi(x)$ when $Z_{i} \sim Exp(\alpha)$, $a = 0.1$, $b = 1.4$, $c = 0.4$, $\lambda = 1$, $x^{*} = 1$ for $\alpha = 0.8, 1, 1.5, 2$.}
	\label{TheTrappingTime-Section3-Figure1}
\end{figure}

\begin{remark}

    Figure \ref{TheTrappingTime-Section3-Figure1-a} shows that the Laplace transform of the trapping time \eqref{TheTrappingTime-Section3-Equation3} approaches the trapping probability as $\delta$ tends to zero, since
    
    \vspace{0.3cm}
    
    \begin{align}
        \lim _{\delta \downarrow 0} m_{\delta}(x) =\mathbb{P}(\tau_{x}<\infty)\equiv\psi(x).
        \label{TheTrappingTime-Section3-Equation4}
    \end{align}
    
    \vspace{0.3cm}
    
    As $\delta\to 0$, \eqref{TheTrappingTime-Section3-Equation3} yields 
    
    \vspace{0.3cm}
    
    \begin{align}
        \psi(x) = \frac{1}{U\left(1-\frac{\lambda}{r},1-\frac{\lambda}{r};0\right)}e^{y(x)}U\left(1-\frac{\lambda}{r},1-\frac{\lambda}{r};-y(x)\right).
        \label{TheTrappingTime-Section3-Equation5}
    \end{align}

    \vspace{0.3cm}
 
We can further simplify the expression for the trapping probability using the  upper incomplete gamma function $\Gamma(a;z)=\int_{z}^{\infty}e^{-t}t^{a-1}dt$. Applying the relation
    
    \vspace{0.3cm}
    
    \begin{align}
        \Gamma(a;z)=e^{-z}U(1-a,1-a;z)
        \label{TheTrappingTime-Section3-Equation6}
    \end{align}
    
    \vspace{0.3cm}
    
    (see equation (13.6.28) of \cite{Book:Abramowitz1964}) and the fact that $\Gamma(a; 0)=\Gamma(a)$ for $\mathbb{R}(a)>0$, we have
    
    \vspace{0.3cm}
    
    \begin{align}
        \psi(x)=\frac{\Gamma\left(\frac{\lambda}{r}; -y(x)\right)}{\Gamma\left(\frac{\lambda}{r}\right)}.
        \label{TheTrappingTime-Section3-Equation7}
    \end{align}
    
    \vspace{0.3cm}

    Figure \ref{TheTrappingTime-Section3-Figure1-b} displays the trapping probability $\psi(x)$ for the stochastic capital process $X_{t}$. Clearly, increasing the value of the parameter $\alpha$ of the exponential distribution of the capital losses reduces the trapping probability for all households, since losses will more likely take values close to zero and will therefore have less impact on households\rq \ capital. 

\end{remark}

\vspace{0.3cm}

\begin{remark}

As an application of the Laplace transform of the trapping time, one quantity of interest is the expected trapping time, i.e. the expected time at which a household will fall into the area of poverty. Reducing a household's trapping probability is central to poverty alleviation. However, knowledge of the time at which a household is expected to fall below the poverty line would allow insurers and governments to better prepare for the potential need to lift a household out of poverty. It provides an alternative comparative measure for the performance analysis of different schemes, helping to inform insurance product design and financial education for consumers. For example, a household with a low expected trapping time may be encouraged to adopt certain risk mitigating behaviours to reduce the impact of shock events and hence the likelihood of them falling below the poverty line. This quantity can be obtained by taking the derivative of $m_{\delta}(x)$, such that
    
    \vspace{0.3cm}

    \begin{align}
        \mathbb{E}\left[\tau_{x} ;\tau_{x}<\infty \right]=-\left.\frac{d}{d \delta} m_{\delta}(x)\right|_{\delta=0},
        \label{TheTrappingTime-Section3-Equation8}
    \end{align}
    
    \vspace{0.3cm}

    where $\mathbb{E}\left[\tau_{x} ;\tau_{x}<\infty \right]$ is analogous to $\mathbb{E}\left[\tau_{x} \mathbbm{1}_{\{\tau_{x}<\infty\}} \right]$.
    
    \end{remark}
   
 \begin{corollary} \label{TheTrappingTime-Section3-Corollary1}
    
    The expected trapping time under the household capital model proposed in  Definition \ref{TheCapitalModel-Section2-Definition1} with initial capital $x \geq x^{*}$, capital growth rate $r$, intensity $\lambda > 0$ and exponentially distributed capital losses with parameter $\alpha > 0$ is given by

    \vspace{0.3cm}
    
    \begin{align}
        \begin{split}
        \mathbb{E}\left[\tau_{x} ;\tau_{x}<\infty \right]&=
        \frac{\Gamma\left(\frac{\lambda}{r};-y(x)\right)}{\lambda U\left(1-\frac{\lambda}{r},1-\frac{\lambda}{r};0\right)}-\frac{\Gamma\left(\frac{\lambda}{r};-y(x)\right)U^{(c)}\left(1-\frac{\lambda}{r},1-\frac{\lambda}{r};0\right)}{r\left[U\left(1-\frac{\lambda}{r},1-\frac{\lambda}{r};0\right)\right]^{2}}\\
        &+e^{y(x)}\frac{U^{(c)}\left(1-\frac{\lambda}{r},1-\frac{\lambda}{r};-y(x)\right)}{rU\left(1-\frac{\lambda}{r},1-\frac{\lambda}{r};0\right)},
        \end{split}
        \label{TheTrappingTime-Section3-Equation9}
    \end{align}
    
    \vspace{0.3cm}    
    
    where $y(x)=-\alpha (x-x^{*})$, $U(\cdot)$ is Tricomi\rq s Confluent Hypergeometric Function as defined in \eqref{Appendix A: Mathematical Proofs-Equation8} and $U^{(c)}(\cdot)$ its derivative with respect to the second parameter as presented in \eqref{Appendix A: Mathematical Proofs-Equation16}.
    
    \end{corollary}

    \vspace{0.3cm}

The mathematical proof of Corollary \ref{TheTrappingTime-Section3-Corollary1} is presented in Appendix \ref{ProofofCorollary3.1}. Note that, the expected trapping time given that trapping occurs can be calculated by taking the following ratio (see for example, equation (4.37) of \cite{Article:Gerber1998}),
    
        \vspace{0.3cm}
    
    \begin{align}
        \begin{split}
        \mathbb{E}\left[\tau_{x}|\tau_{x}<\infty\right]= \frac{\mathbb{E}\left[\tau_{x};\tau_{x}<\infty\right]}{\psi(x)}.
        \end{split}
        \label{TheTrappingTime-Section3-Equation10}
    \end{align}
    
    \vspace{0.3cm}  
    
\begin{figure}[H]
        \centering
	    \includegraphics[width=8cm, height=8cm]{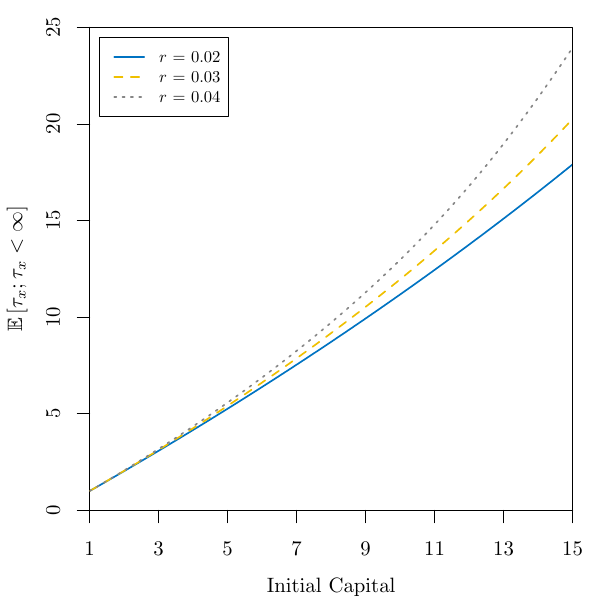}
	    \caption{Expected trapping time when $Z_{i} \sim Exp(1)$, $\lambda = 1$ and $x^{*} = 1$ for $r = 0.02,0.03,0.04$.}
	    \label{TheTrappingTime-Section3-Figure2}
    \end{figure}

    In line with intuition, the expected trapping time is an increasing function of both the capital growth rate $r$ and initial capital $x$. A number of expected trapping times for varying values of $r$ are displayed in Figure \ref{TheTrappingTime-Section3-Figure2}.

    \vspace{0.3cm}

\begin{remark}

The  ruin probability for the insurance model \eqref{TheCapitalModel-Section2-Equation5} given by

\vspace{0.3cm}

\begin{align}
    \xi (u)= \mathbb{P}(U_t<0 \text{ for some }t>0 \mid U_0=u)
    \label{TheTrappingTime-Section3-Equation11},
\end{align}

\vspace{0.3cm}

is found by \cite{Article:Sundt1995} to satisfy the Integro-Differential Equation (IDE)

\vspace{0.3cm}

\begin{align}
    (\nu u+p)\xi'(u)-\lambda \xi (u)+ \lambda \int_0^ {u} \xi(u-z) \, dG_{Z}(z)+\lambda(1-G_{Z}(u))=0, \qquad u\geq 0.
    \label{TheTrappingTime-Section3-Equation12}
\end{align}

\vspace{0.3cm}

Note that when $p=0$, 
\eqref{TheTrappingTime-Section3-Equation12} coincides with the special case of
\eqref{Appendix A: Mathematical Proofs-Equation1} when $x^*=0$, $w(x_{1},x_{2})=1$ and $\delta=0$. Thus, the household trapping time can be thought of as the insurer\rq s ruin time. Indeed, the ruin probability in the case of exponential claims when $p=0$, as shown in Section 6 of \cite{Article:Sundt1995}, is exactly the same as the trapping probability \eqref{TheTrappingTime-Section3-Equation7} when $x^*=0$.

\end{remark}

\section{Introducing Microinsurance}\label{IntroducingMicroinsurance-Section4}

As in \cite{Article:Kovacevic2011}, we assume households have the option of enrolling in a microinsurance scheme that covers a certain proportion of the capital losses they encounter. The microinsurance policy has proportionality factor $1-\kappa$, where $\kappa \in [0,1]$, such that $100 \cdot (1-\kappa)$ percent of the damage is covered by the microinsurance provider. The premium rate paid by households, calculated according to the expected value principle, is given by

\vspace{0.3cm}

\begin{align}
    \pi(\kappa, \theta)=(1+\theta) \cdot(1-\kappa) \cdot \lambda \cdot \mathbb{E}\left[Z_{i}\right], \label{IntroducingMicroinsurance-Section4-Equation1}
\end{align}

\vspace{0.3cm}

where $\theta$ is some loading factor. The expected value principle is popular due to its simplicity and transparency. When $\theta = 0$, one can consider $\pi(\kappa, \theta)$ to be the pure risk premium \citep{Book:Albrecher2017}.

The stochastic capital process of a household covered by a microinsurance policy is denoted by $X_{t}^{\scaleto{(\kappa)}{5pt}}$. We differentiate between all variables and parameters relating to the original uninsured and the insured processes through use of the superscript $(\kappa)$ in the latter case. We assume the basic model parameters are unchanged by the introduction of microinsurance coverage (parameters $a, b$ and $c$ of \cite{Article:Kovacevic2011}, previously introduced in Section \ref{TheCapitalModel-Section2}). Here, we only allow households to select a fixed retention rate, while other studies look for an optimal retention rate process that maximises the expected discounted capital by admitting adjustments in the retention rate after each capital loss throughout the lifetime of the insurance contract (see, for example, \cite{Inbook:Kovacevic2021}).

Since premiums are paid from a household's income, the capital growth rate $r$ is adjusted such that it reflects the lower rate of income generation resulting from the need for premium payment. The premium rate is restricted to prevent certain poverty, which would occur should it exceed the rate of income generation. The capital growth rate of the insured household $r^{(\kappa)}=\left(1-a\right)\cdot \left(b-\pi \right)\cdot c  > 0$ is lower than that of the uninsured household, while the critical capital is higher (see \cite{Article:Kovacevic2011} for further discussion). Note that, previous work such as that of \cite{Article:Janzen2020} allow households to choose optimal levels of consumption and insurance coverage over time based on asset holdings and the probability distribution of future assets. Here, all households who can afford to buy insurance enrol in a scheme; that is, as mentioned above, households whose income generation rate is greater than the insurance premium are able to choose any affordable insurance coverage, therefore admitting both optimal and suboptimal choices with respect to the trapping probability. Although this feature aligns with the low levels of financial literacy that characterise the microinsurance environment \citep{Book:Churchill2012}, it could initially be considered as a limitation of our model. However, one of the core objectives of the subsidised schemes introduced in Sections \ref{MicroinsurancewithSubsidisedConstantPremiums-Section5} and \ref{MicroinsurancewithSubsidisedFlexiblePremiums-Section6} is to diminish the adverse effects that arise with suboptimal choices and as such any limitation is accounted for.

In between jumps, the insured stochastic growth process $X_{t}^{\scaleto{(\kappa)}{5pt}}$ behaves in the same manner as \eqref{TheCapitalModel-Section2-Equation2}, with parameters corresponding to the proportional insurance case of this section. By enrolling in a microinsurance scheme a household\rq s capital losses are reduced  to $Y_{i} :=\kappa \cdot Z_{i}$. Considering the case in which losses follow an exponential distribution with parameter $\alpha > 0$, the structure of the IDE \eqref{Appendix A: Mathematical Proofs-Equation1} remains the same. However, acquisition of a proportional microinsurance policy changes the parameter of the distribution $G_{Y}$ of the random losses ($Y_i$). Namely, we have that $Y_{i} \sim Exp\left(\alpha^{\scaleto{(\kappa)}{5pt}}\right)$ for $\kappa \in (0,1]$, where $\alpha^{\scaleto{(\kappa)}{5pt}} := \frac{\alpha}{\kappa}$. 

Following a similar procedure to that in Proposition \ref{TheTrappingTime-Section3-Proposition1} (as presented in Appendix \ref{Appendix A: Mathematical Proofs}), one easily obtains the Laplace transform of the trapping time and thus the trapping probability for the insured process.

\vspace{0.3cm}

\begin{proposition}\label{IntroducingMicroinsurance-Section4-Proposition1}

Consider the capital process of a household enrolled in a microinsurance scheme with proportionality factor $1-\kappa \in [0,1]$ (as introduced in this section). Assume the household has initial capital $x\ge x^{\scaleto{(\kappa)*}{5pt}}$, capital growth rate $r^{\scaleto{(\kappa)}{5pt}}$, intensity $\lambda > 0$ and exponentially distributed capital losses with parameter $\alpha^{\scaleto{(\kappa)}{5pt}} > 0$. The Laplace transform of the trapping time is given by

\vspace{0.3cm}

\begin{align}
    m_{\delta}^{\scaleto{(\kappa)}{5pt}}(x)=\frac{\lambda}{(\lambda + \delta) U\left(1-\frac{\lambda}{r^{\scaleto{(\kappa)}{5pt}}},1-\frac{\lambda+\delta}{r^{\scaleto{(\kappa)}{5pt}}};0\right)}e^{y^{\scaleto{(\kappa)}{5pt}}(x)}U\left(1-\frac{\lambda}{r^{\scaleto{(\kappa)}{5pt}}},1-\frac{\lambda+\delta}{r^{\scaleto{(\kappa)}{5pt}}};-y^{\scaleto{(\kappa)}{5pt}}(x)\right),
    \label{IntroducingMicroinsurance-Section4-Equation2}
 \end{align}
 
\vspace{0.3cm}
 
where $\delta \ge 0$ is the force of interest for valuation and $y^{\scaleto{(\kappa)}{5pt}}(x)=-\alpha^{\scaleto{(\kappa)}{5pt}}\left(x-x^{\scaleto{(\kappa)*}{5pt}}\right)$. 

\end{proposition}

\begin{figure}[H]
	\begin{subfigure}[b]{0.5\linewidth}
  		\includegraphics[width=8cm, height=8cm]{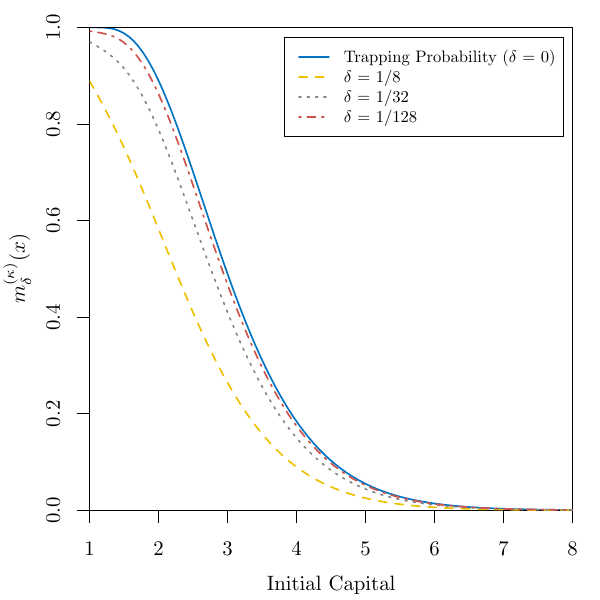}
		\caption{}
  		\label{IntroducingMicroinsurance-Section4-Figure1-a}
	\end{subfigure}
	\begin{subfigure}[b]{0.5\linewidth}
  		\includegraphics[width=8cm, height=8cm]{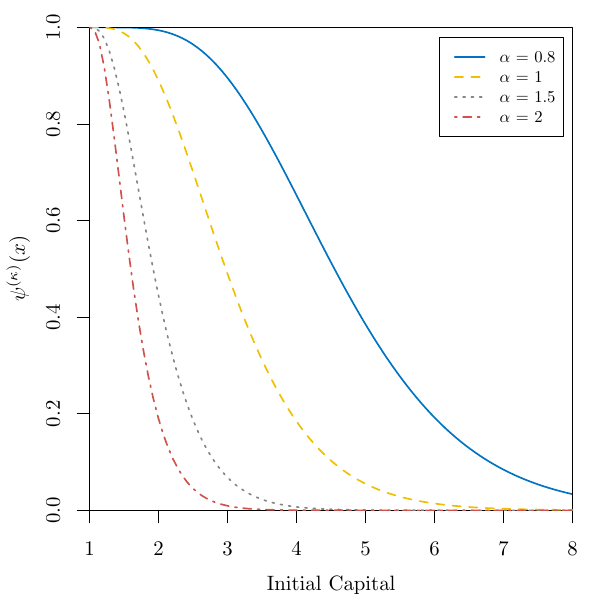}
		\caption{}
  		\label{IntroducingMicroinsurance-Section4-Figure1-b}
	\end{subfigure}
	\caption{(a) Laplace transform $m_{\delta}^{\scaleto{(\kappa)}{5pt}}(x)$ of the trapping time when $Z_{i} \sim Exp(1)$, $a = 0.1$, $b = 1.4$, $c = 0.4$, $\lambda = 1$, $x^{\scaleto{(\kappa)*}{5pt}} = 1$, $\kappa = 0.5$ and $\theta=0.5$ for $\delta = 0, \frac{1}{8}, \frac{1}{32}, \frac{1}{128}$ (b) Trapping probability $\psi^{\scaleto{(\kappa)}{5pt}}(x)$ when $Z_{i} \sim Exp(\alpha)$, $a = 0.1$, $b = 1.4$, $c = 0.4$, $\lambda = 1$, $x^{\scaleto{(\kappa)*}{5pt}} = 1$, $\kappa = 0.5$ and $\theta=0.5$ for $\alpha = 0.8, 1, 1.5, 2$.}
	\label{IntroducingMicroinsurance-Section4-Figure1}
\end{figure}

Figure \ref{IntroducingMicroinsurance-Section4-Figure1-a} displays the Laplace transform $m_{\delta}^{\scaleto{(\kappa)}{5pt}}(x)$ for varying values of $\delta$. As mentioned earlier, as $\delta \rightarrow 0$, the Laplace transform $m_{\delta}^{\scaleto{(\kappa)}{5pt}}(x)$ converges to the trapping probability $\psi^{\scaleto{(\kappa)}{5pt}}(x)$.

\begin{remark}

    The trapping probability of the insured process $\psi^{\scaleto{(\kappa)}{5pt}}(x)$, displayed in Figure \ref{IntroducingMicroinsurance-Section4-Figure1-b}, is given by
    
\vspace{0.3cm}
    
\begin{align}
    \psi^{\scaleto{(\kappa)}{5pt}}(x)=\frac{\Gamma\left(\frac{\lambda}{r^{\scaleto{(\kappa)}{5pt}}}; -y^{\scaleto{(\kappa)}{5pt}}(x)\right)}{\Gamma\left(\frac{\lambda}{r^{\scaleto{(\kappa)}{5pt}}}\right)}.
    \label{IntroducingMicroinsurance-Section4-Equation3}
\end{align}

As mentioned previously, increasing the value of the parameter $\alpha$ of the exponential distribution of the capital losses reduces the trapping probability since capital losses are likely to have a low impact on household capital. Moreover, note that as the proportionality factor $\kappa \rightarrow 0$, the parameter $\alpha^{\scaleto{(\kappa)}{5pt}} := \frac{\alpha}{\kappa}$ of the new capital losses $Y_{i}$ of the insured capital process increases, leading households to experience low impact losses with a higher probability ($Y_{i}$ will likely have values close to zero), but to pay higher premiums \eqref{IntroducingMicroinsurance-Section4-Equation1}.

\vspace{0.3cm}

\end{remark}
     
    \begin{remark}
    When $\kappa = 0$ the household has full microinsurance coverage, as the microinsurance provider covers the total capital loss experienced by the household. On the other hand, when $\kappa = 1$, no coverage is provided by the insurer, i.e. $X_{t}=X_{t}^{\scaleto{(\kappa)}{5pt}}$.
    \end{remark}

\begin{figure}[H]
    \centering
	\includegraphics[width=8cm, height=8cm]{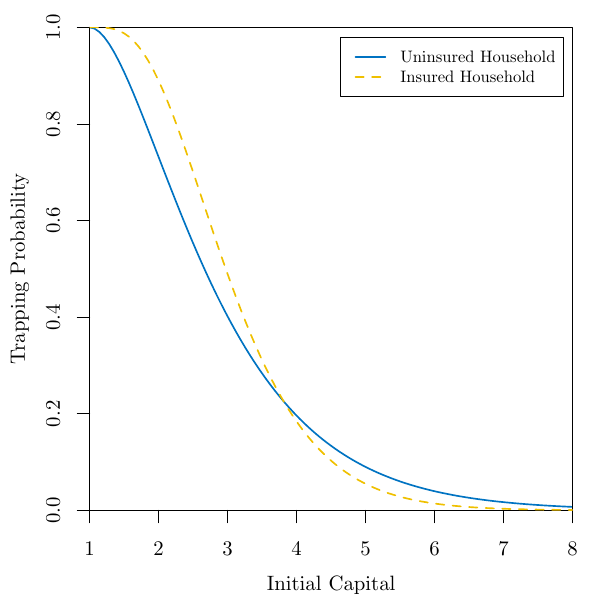}
	\caption{Trapping probabilities for the uninsured and insured capital processes when $Z_{i} \sim Exp(1)$, $a = 0.1$, $b = 1.4$, $c = 0.4$, $\lambda = 1$, $\kappa = 0.5$, $\theta=0.5$ and $x^{*} = x^{\scaleto{(\kappa)*}{5pt}} = 1$.}
	\label{IntroducingMicroinsurance-Section4-Figure2}
\end{figure}

Figure \ref{IntroducingMicroinsurance-Section4-Figure2} presents a comparison between the trapping probabilities of the insured and uninsured processes. As in \cite{Article:Kovacevic2011}, households with initial capital close to the critical capital (here, the critical capital $x^{*}$ is 1), i.e. the most vulnerable households, may not receive a real benefit from enrolling in a microinsurance scheme. Although subscribing to a proportional microinsurance scheme reduces capital losses, premium payments appear to make such households more prone to falling into the area of poverty. The intersection point of the two probabilities in Figure \ref{IntroducingMicroinsurance-Section4-Figure2} corresponds to the boundary between households that benefit from the uptake of microinsurance and those who are adversely affected. 

\section{Microinsurance with Subsidised Constant Premiums}\label{MicroinsurancewithSubsidisedConstantPremiums-Section5}

\subsection{General Setting}\label{GeneralSetting-Subection51}

\begin{figure}[H]
	\begin{subfigure}[b]{0.5\linewidth}
  		\includegraphics[width=8cm, height=8cm]{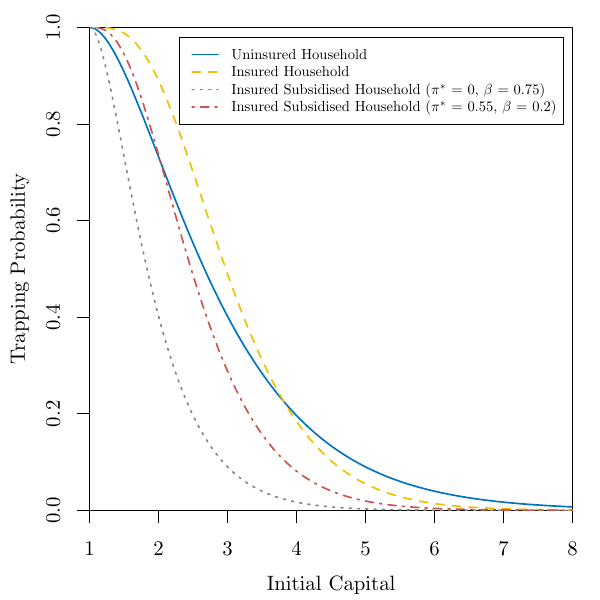}
		\caption{}
  		\label{GeneralSetting-Subection51-Figure1-a}
	\end{subfigure}
	\begin{subfigure}[b]{0.5\linewidth}
  		\includegraphics[width=8cm, height=8cm]{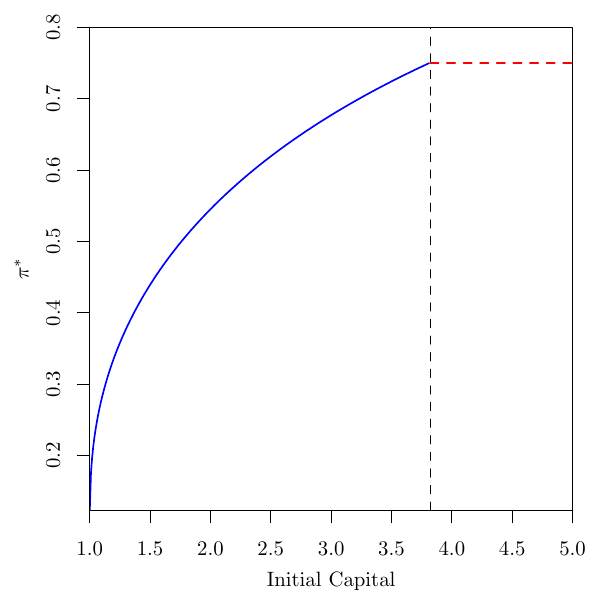}
		\caption{}
  		\label{GeneralSetting-Subection51-Figure1-b}
	\end{subfigure}
	\caption{(a) Trapping probabilities for the uninsured, insured and insured subsidised capital processes when $Z_{i} \sim Exp(1)$, $a = 0.1$, $b = 1.4$, $c = 0.4$, $\lambda = 1$, $x^{*} = x^{\scaleto{(\kappa)*}{5pt}} = x^{\scaleto{\pi(\kappa, \theta)*}{5pt}} = 1$, $\kappa = 0.5$, $\theta = 0.5$ and $\pi = 0.75$ for $\pi^{*} = 0, 0.55$ (b) Optimal $\pi^{*}$ for varying initial capital when $Z_{i} \sim Exp(1)$, $a = 0.1$, $b = 1.4$, $c = 0.4$, $\lambda = 1$, $x^{\scaleto{\pi(\kappa, \theta)*}{5pt}} = 1$, $\kappa = 0.5$, $\theta = 0.5$ and $\pi = 0.75$.}
	\label{GeneralSetting-Subection51-Figure1}
\end{figure}

The preliminary results suggest that microinsurance alone may not be enough to reduce the likelihood of impoverishment for those closest to the poverty line, and so additional aid is required. In this section, we study the cost-effectiveness of government subsidised premiums, considering the case in which the government subsidises an amount $\beta = \pi - \pi^{*}$, where $\pi^{*}$ is the premium after subsidisation, such that $\pi \geq \pi^{*} \geq 0$. As such, lower values of $\pi^{*}$ correspond to greater government support. When $\pi^{*} = 0$ the premium is fully subsidised, whereas when $\pi = \pi^{*}$ households do not receive any subsidies. Note that, in contrast to previous work such as that of \cite{Article:Kovacevic2011}, where the subsidy is limited to the loading factor, and the self-targeted subsidy strategy of \cite{Article:Janzen2020}, where fixed subsidies are provided uniformly to poor households who would anyway purchase insurance, here we extend the possibility of households benefiting from greater subsidisation in line with existing government supported microinsurance schemes, while adjusting for the governmental cost. Some examples for varying levels of subsidisation are presented in Figure \ref{GeneralSetting-Subection51-Figure1-a}.

Naturally, we assume that governments are interested in optimising the subsidy provided to households. Governments should provide subsidies to microinsurance providers such that they enhance households\rq \,  benefits of enrolling in microinsurance schemes, however, they also need to gauge the cost-effectiveness of subsidy provision. Households with capital very close to the critical capital will require additional aid, while government support is not necessarily essential for more privileged households. Since all non-zero values of $\pi^{*}$ below the optimal value will induce a trapping probability lower than that of the uninsured process through a reduction in premium, one approach to determining the optimal subsidy for households that require government aid is to find the solution of the equation

\vspace{0.3cm}

\begin{align}
    \psi^{\scaleto{\pi^{*}(\kappa,\theta)}{5pt}}(x)=\psi(x),
    \label{GeneralSetting-Subection51-Equation1}
\end{align}

\vspace{0.3cm}

where $\psi^{\scaleto{\pi^{*}(\kappa,\theta)}{5pt}}(x)$ and $\psi(x)$ denote the trapping probabilities of the insured subsidised and uninsured processes, respectively. The behaviours of these trapping probabilities can be seen in Figure \ref{GeneralSetting-Subection51-Figure1-a}, while the most privileged households do not need help from the government since the non-subsidised insurance lowers their trapping probability below the uninsured, the poorest individuals may require further support. Figure \ref{GeneralSetting-Subection51-Figure1-b} illustrates the optimal value of $\pi^{*}$ for varying initial capital, verifying that, from the point at which the yellow dashed line (insured household) intersects the blue solid line (uninsured household) in Figure \ref{GeneralSetting-Subection51-Figure1-a}, payment of the entire premium is affordable for the most privileged households, with the optimal premium remaining constant at $\pi^{*}=\pi=0.75$ after this point (red dashed line in Figure \ref{GeneralSetting-Subection51-Figure1-b}).

\subsection{Cost of Social Protection}\label{CostofSocialProtection-Subsection52}

Next, we assess government cost-effectiveness of the provision of microinsurance premium subsidies to households. Let $\tau^{\scaleto{\pi^{*}(\kappa,\theta)}{5pt}}_{x}$ denote the trapping time of a household covered by a subsidised microinsurance policy. Moreover, let $\delta \geq 0$ be the force of interest for valuation and let $S$ denote the present value of all subsidies provided by government until the trapping time such that

\vspace{0.3cm}

\begin{align}
    S=\beta\int^{\tau^{\scaleto{\pi^{*}(\kappa,\theta)}{5pt}}_{x}}_{0} e^{-\delta t} dt =\beta \ax*{\angl{\tau^{\scaleto{\pi^{*}(\kappa,\theta)}{5pt}}_{x}}}. \label{CostofSocialProtection-Subsection52-Equation1}
\end{align}

\vspace{0.3cm}

We assume governments provide subsidies according to the strategy introduced in Section \ref{GeneralSetting-Subection51}, i.e. the government subsidises an amount $\beta = \pi - \pi^{*}$.

For $x\geq x^{\scaleto{\pi^{*}(\kappa,\theta)*}{5pt}}$, where $x^{\scaleto{\pi^{*}(\kappa,\theta)*}{5pt}}$ denotes the critical capital of the insured subsidised process, let $V^{\scaleto{\pi^{*}(\kappa,\theta)}{5pt}}(x)$ be the expected discounted premium subsidies provided by the government to a household with initial capital $x$ until the trapping time, that is, 

\vspace{0.3cm}

\begin{align}
     V^{\scaleto{\pi^{*}(\kappa,\theta)}{5pt}}(x)= \mathbb{E}\left[S \mid X^{\scaleto{\pi^{*}(\kappa,\theta)}{5pt}}_{0}=x\right].
     \label{CostofSocialProtection-Subsection52-Equation2}
\end{align}

\vspace{0.3cm}
 
 \begin{proposition}\label{CostofSocialProtection-Subsection52-Proposition1}
 
Consider a household enrolled in a microinsurance scheme with subsidised constant premiums in which the government subsidises an amount $\beta=\pi - \pi^{*}$, where $\pi \geq \pi^{*} \geq 0$ (as discussed in Section \ref{GeneralSetting-Subection51}), with proportionality factor $1-\kappa \in [0,1]$. Assume an initial capital $x \ge x^{\scaleto{\pi^{*}(\kappa,\theta)*}{5pt}}$, capital growth rate $r^{\scaleto{\pi^{*}(\kappa,\theta)}{5pt}}$, intensity $\lambda > 0$ and exponentially distributed capital losses with parameter $\alpha^{\scaleto{(\kappa)}{5pt}} > 0$. The expected discounted premium subsidies provided by the government to the household until the trapping time is given by
 
 \vspace{0.3cm}
 
\begin{equation}
\begin{aligned}
     V^{\scaleto{\pi^{*}(\kappa,\theta)}{5pt}}(x)&= \frac{\beta}{\delta}\left[1-m_{\delta}^{\scaleto{\pi^{*}(\kappa,\theta)}{5pt}}(x)\right],
     \end{aligned}
     \label{CostofSocialProtection-Subsection52-Equation3}
\end{equation}

\vspace{0.3cm}
 
where $m^{\scaleto{\pi^{*}(\kappa,\theta)}{5pt}}_\delta(x)$ is the Laplace transform of the trapping time with rate $r^{\scaleto{\pi^{*}(\kappa,\theta)}{5pt}}$ and critical capital $x^{\scaleto{\pi^{*}(\kappa,\theta)*}{5pt}}$.
 
 \end{proposition}

\vspace{0.3cm}

See Appendix \ref{ProofofProposition5.1} for proof of Proposition \ref{CostofSocialProtection-Subsection52-Proposition1}. We now formally define the government\rq s cost of social protection.
 
\vspace{0.3cm}

\begin{definition}\label{TheTrappingTime-Section3-Definition1}

Consider \eqref{TheTrappingTime-Section3-Equation2}, the expected discounted penalty function at trapping of a household enrolled in a subsidised microinsurance scheme with initial capital $x$. Let $w(x_{1},x_{2})= x_{2} + M^{\scaleto{(\kappa)}{5pt}}-x^{\scaleto{\pi^{*}(\kappa,\theta)*}{5pt}}$ be the penalty function, where $M^{\scaleto{(\kappa)}{5pt}} \geq x^{\scaleto{\pi^{*}(\kappa,\theta)*}{5pt}}$. Here, $M^{\scaleto{(\kappa)}{5pt}} - x^{\scaleto{\pi^{*}(\kappa,\theta)*}{5pt}}$ is a constant representing the cost to lift households further away from the area of poverty, while $x_{2}$ accounts for the cost to lift households up to the critical level $x^{\scaleto{\pi^{*}(\kappa,\theta)*}{5pt}}$. Thus, the expected discounted penalty function at trapping $m^{\scaleto{\pi^{*}(\kappa,\theta)}{5pt}}_{\delta,w}(x)$ is the expected present value of the capital deficit at trapping plus a cost $M^{\scaleto{(\kappa)}{5pt}}-x^{\scaleto{\pi^{*}(\kappa,\theta)*}{5pt}}$ due at the trapping time. We therefore define a government\rq s cost of social protection as the expected discounted premium subsidies, given by \eqref{CostofSocialProtection-Subsection52-Equation3}, plus the expected present value of the capital deficit and the amount $M^{\scaleto{(\kappa)}{5pt}}-x^{\scaleto{\pi^{*}(\kappa,\theta)*}{5pt}}$. 

\end{definition}

\vspace{0.3cm}

\begin{figure}[H]
    \centering
	\includegraphics[width=8cm, height=8cm]{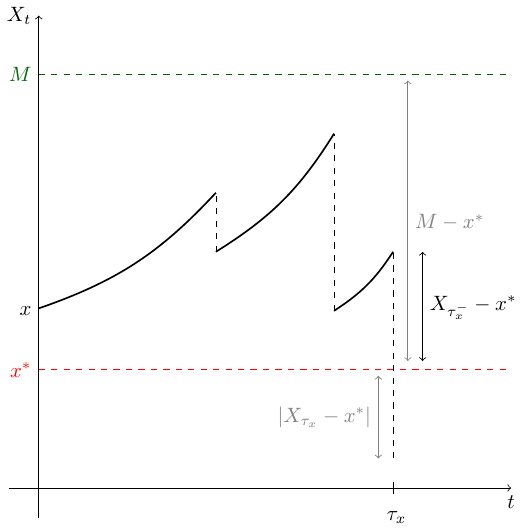}
	\caption{The cost incurred by the government at the trapping time is given by $\mid X_{\tau_{x}} -x^{*}\mid$, the capital deficit at the trapping time, plus $M - x^{*}$, the cost to lift households further away from the area of poverty.}
	\label{CostofSocialProtection-Subsection52-Figure1}
\end{figure}

\begin{remark}

The government does not provide subsidies for uninsured households. We consider their expected discounted penalty function at trapping to be $m_{\delta,w}(x)$ with $w(x_{1},x_{2})= x_{2} + M - x^{*}$. The choice of this particular penalty function is based on the idea that the government, in order to lift a household out of poverty, incurs a cost equal to the household\rq s capital deficit at the moment they fall into poverty plus a fixed cost $M - x^{*}$ that ensures, with a certain level of confidence, that the household will not return to poverty. This differs from other approaches taken in previous research in which the cost of social protection considers only the present value of the transfers needed to close the poverty gap (see, for example, \cite{Inbook:Barrett2016} and \cite{Article:Janzen2020}). In this way, the likelihood of re-incurring these costs for the same household is reduced. Thus, the constant $M$ could be defined in such a way that the government ensures with some probability that households will not fall into the area of poverty again. For instance, let us consider $\epsilon$ to be the most admissible trapping probability for an uninsured household. We can therefore define the statistic

\vspace{0.3cm}

\begin{align}
    M:=\inf \left\{x \geq x^{*}: \psi(x)<\epsilon\right\},
    \label{CostofSocialProtection-Subsection52-Equation4}
\end{align}

\vspace{0.3cm}

where $M$ is the Minimum Initial Capital (MIC) required to ensure a trapping probability of less than $\epsilon$. This statistic has also been studied from the point of view of an insurance company, where $\epsilon$ represents the most admissible probability that the insurance company will become insolvent \citep{Article:Sattayatham2013, Article:Constantinescu2019}. As such, the government determines an appropriate amount $M$ such that a household's probability of re-entering the area of poverty is less than $\epsilon \in (0,1)$. Clearly, higher values of $M$ will increase the certainty that households will not return to poverty. Also note that the value of $M$ will differ between uninsured, insured and insured subsidised households due to the fact that their trapping probabilities are distinct. However, in this study we assume that governments will consider an amount $M^{\scaleto{(\kappa)}{5pt}}$ under all microinsurance schemes (with or without subsidies). That is, we assume that households who are initially enrolled in a microinsurance scheme (with or without subsidies) will be enrolled in a scheme without subsidies just after being lifted away from the area of poverty. Figure \ref{CostofSocialProtection-Subsection52-Figure1} displays the cost incurred at the trapping time when employing the penalty function $w(x_{1},x_{2})= x_{2} + M - x^{*}$ for an uninsured household. 

\end{remark}

\begin{remark}

The government manages selection of an appropriate force of interest $\delta \geq 0$. For lower force of interest the government discounts future subsidies more heavily, while for higher interest future subsidies almost vanish.
\end{remark}

\vspace{0.3cm}

\begin{remark}

When losses are exponentially distributed with parameter $\alpha^{\scaleto{(\kappa)}{5pt}} >0$, one can obtain a closed form expression for the cost of social protection. Given our derivation of $V^{\scaleto{\pi^{*}(\kappa,\theta)}{5pt}}(x)$ in \eqref{CostofSocialProtection-Subsection52-Equation3}, we determine an expression for $m_{\delta,w}^{\scaleto{\pi^{*}(\kappa,\theta)}{5pt}}(x)$.
\end{remark}

\vspace{0.3cm}

\begin{proposition}\label{CostofSocialProtection-Subsection52-Proposition2}

Consider a household enrolled in a microinsurance scheme with subsidised constant premiums in which the government subsidises an amount $\beta=\pi - \pi^{*}$, where $\pi \geq \pi^{*} \geq 0$, with proportionality factor $1-\kappa \in [0,1]$. Assume an initial capital $x \ge x^{\scaleto{\pi^{*}(\kappa,\theta)*}{5pt}}$, capital growth rate $r^{\scaleto{\pi^{*}(\kappa,\theta)}{5pt}}$, intensity $\lambda > 0$ and exponentially distributed capital losses with parameter $\alpha^{\scaleto{(\kappa)}{5pt}} > 0$. Furthermore, let $M^{\scaleto{(\kappa)}{5pt}} - x^{\scaleto{\pi^{*}(\kappa,\theta)*}{5pt}}$, with $M^{\scaleto{(\kappa)}{5pt}} \geq x^{\scaleto{\pi^{*}(\kappa,\theta)*}{5pt}}$, be the cost to lift households further away from the area of poverty. The expected discounted cost incurred by the government at the trapping time is given by

\vspace{0.3cm}
{
\begin{equation}
\begin{aligned}
m^{\scaleto{\pi^{*}(\kappa,\theta)}{5pt}}_{\delta,w}(x) = \left[\frac{1}{\alpha^{\scaleto{(\kappa)}{5pt}}}+ M^{\scaleto{(\kappa)}{5pt}}-x^{\scaleto{\pi^{*}(\kappa,\theta)*}{5pt}}\right] m^{\scaleto{\pi^{*}(\kappa,\theta)}{5pt}}_{\delta}(x),\label{CostofSocialProtection-Subsection52-Equation5}
\end{aligned}
\end{equation}
}
\vspace{0.3cm}

where $m^{\scaleto{\pi^{*}(\kappa,\theta)}{5pt}}_\delta(x)$ is the Laplace transform of the trapping time with rate $r^{\scaleto{\pi^{*}(\kappa,\theta)}{5pt}}$ and critical capital $x^{\scaleto{\pi^{*}(\kappa,\theta)*}{5pt}}$, and $\delta \ge 0$ is the force of interest for valuation.

\end{proposition}

\vspace{0.3cm}

See Appendix \ref{ProofofProposition5.2} for proof of Proposition \ref{CostofSocialProtection-Subsection52-Proposition2}.

\begin{figure}[H]
    \centering
	\includegraphics[width=8cm, height=8cm]{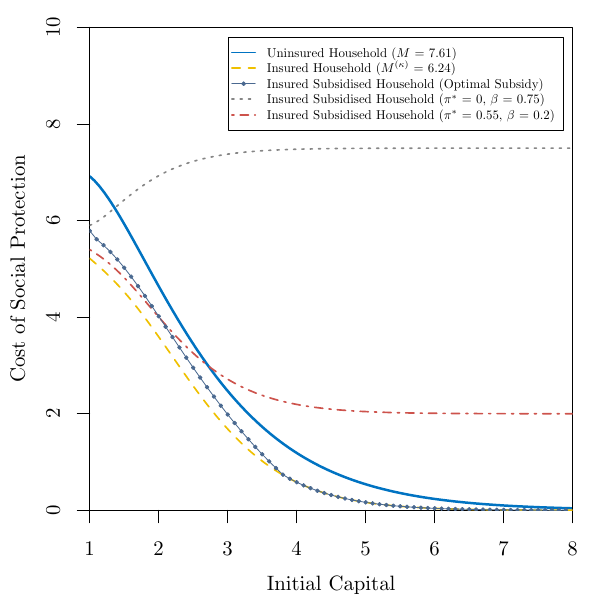}
	\caption{Cost of social protection for the uninsured, insured and insured subsidised with $\pi^{*}= \pi^{*}_{\scaleto{\text{ {\fontfamily{qcr}\selectfont
 Optimal}}}{4pt}}, 0, 0.55$ capital processes when $Z_{i}\sim Exp(1)$, $a = 0.1$, $b = 1.4$, $c = 0.4$, $\lambda=1$, $x^{*}= x^{\scaleto{(\kappa)*}{5pt}} = x^{\scaleto{\pi(\kappa, \theta)*}{5pt}} = 1$, $\kappa = 0.5$, $\theta = 0.5$, $\delta =0.1$, $\epsilon = 0.01$ and $\pi= 0.75$.}
	\label{CostofSocialProtection-Subsection52-Figure2}
\end{figure}

\begin{remark}

Due to the lack-of-memory property of the exponential distribution the deficit at trapping, given that trapping occurs, is
exponentially distributed. One can easily verify this by specifying the penalty function such that for any fixed $u$, $w(x_{1}, x_{2})=\mathbbm{1}_{\{x_{2} \leq u\}}$ and $\delta = 0$. Similar results to that of Proposition \ref{CostofSocialProtection-Subsection52-Proposition2} have been obtained for other risk processes (see,  for instance, Example 3.2 of \cite{Article:Albrecher2005}). 
\end{remark}

Figure \ref{CostofSocialProtection-Subsection52-Figure2} displays the governmental cost of social protection for varying initial capital. Here, the value of $M$ is given by  \eqref{CostofSocialProtection-Subsection52-Equation4}. Note that, as mentioned previously, high values of $\delta$ hand a lower weight to future government subsidies, whereas high values of $M$ grant greater certainty that a household will not return to the area of poverty once lifted out of it.

Governments do not benefit from subsidising insurance for the most privileged households since they will subsidise premiums indefinitely, almost surely. Hence, as also illustrated in Figure \ref{GeneralSetting-Subection51-Figure1-b}, it is favourable for governments to remove subsidies for this particular household group since their cost of social protection in Figure \ref{CostofSocialProtection-Subsection52-Figure2} (red dashed-dotted and gray dotted lines for highest values of initial capital) is higher than when uninsured (blue solid line for highest values of initial capital). This is largely due to the fact that governments are still obliged to subsidise a given amount of the premium even though greater initial capital leads to lower trapping probabilities and therefore a reduction in the likelihood of the government needing to lift these households away from the area of poverty. 

In addition, Figure \ref{CostofSocialProtection-Subsection52-Figure2} shows that when providing optimal subsidies, governments can reduce the cost of social protection incurred. Here, the fully subsidised scheme (when $\pi^{*}=0$) has a higher cost for all households relative to the scheme that provides optimal subsidies (blue circular-marked line below the gray dotted line), and the difference between the two increases as the initial capital increases, until the moment at which the cost of social protection for the fully subsidised scheme and the scheme that provides optimal subsidies converge to $\frac{\beta}{\delta}$ and zero, respectively. Similarly, for more privileged households, a subsidised scheme (with $\pi^{*}=0.55$) has a higher cost relative to the optimal case (blue circular-marked line below the red dashed-dotted line for more privileged households). On the contrary, for the most vulnerable, Figure \ref{CostofSocialProtection-Subsection52-Figure2} shows the cost of social protection of the scheme that provides optimal subsidies is above the subsidised scheme (blue circular-marked line above the red dashed-dotted line for most vulnerable households) as optimal subsidies for this group provide greater support (i.e. the optimal values for $\pi^{*}$ are lower). Note that the cost of social protection for the insured (yellow dashed line), non-optimal insured subsidised (gray dotted and red dashed-dotted lines) for the most vulnerable (for those with initial capital lower than $x=1.362$ when $\pi^{*}=0$ and $x=2.719$ when $\pi^{*}=0.55$) and optimal insured subsidised households (blue circular-marked line) is below that of the uninsured, thus highlighting the significance of insurance as a tool for reducing the governmental cost of social protection. Moreover, it is not surprising that the cost of social protection for the insured (yellow dashed line), is lower than the cost of social protection for the rest of the microinsurance schemes, since governments will not provide subsidies, but only an injection of capital in the event of entering poverty. 

\section{Microinsurance with Subsidised Flexible Premiums}\label{MicroinsurancewithSubsidisedFlexiblePremiums-Section6}

\subsection{General Setting}\label{GeneralSetting-Subsection61}

Since microinsurance premiums are generally paid as soon as coverage is purchased, a household\rq s capital growth could be constrained after joining a scheme, as observed in the results of Sections \ref{IntroducingMicroinsurance-Section4} and \ref{MicroinsurancewithSubsidisedConstantPremiums-Section5}. It is therefore interesting to consider alternative premium payment mechanisms. From the point of view of microinsurance providers, advance premium payments are preferred so that additional income can be generated through investment, naturally leading to lower premium rates. Consumers on the other hand may find it difficult to pay premiums up front. This is a common problem in low-income populations, with research suggesting that consumers prefer to pay smaller installments over time \citep{Book:Churchill2012}. Collecting premiums at a time that is inconvenient for households can be futile. Alternative insurance designs in which premium payments are delayed until the  insured\rq s income is realised and any indemnities are paid have also been studied. Under such designs, insurance take-up increases, since liquidity constraints are relaxed and concerns regarding insurer default, also prevalent in low-income classes, reduce \citep{Article:Liu2016}.  

In this section, we introduce an alternative microinsurance subsidy scheme with flexible premium payments. We denote the capital process of a household enrolled in the alternative microinsurance subsidy scheme by $X_{t}^{\scaleto{(\mathcal{A})}{5pt}}$. As in Section \ref{IntroducingMicroinsurance-Section4}, we differentiate between variables and parameters relating to the uninsured, insured and alternative insured processes using the superscript $(\mathcal{A})$. Under such an alternative microinsurance subsidy scheme households pay premiums when their capital is above some capital barrier $B \ge x^{\scaleto{(\mathcal{A})*}{5pt}}$, with the premium otherwise paid by the government. In other words, whenever the insured capital process is below the capital level $B$ premiums are entirely subsidised by the government, however, when a household's capital is above $B$ the premium $\pi$ is paid continuously by the household itself. This method of premium collection may motivate households to maintain a level of capital below $B$ in order to avoid premium payments. Consequently, for the purpose of this study, we assume that households always pursue capital growth. Our aim is to study how this alternative microinsurance subsidy scheme could help households reduce their probability of falling into the area of poverty. In addition, we measure the cost-effectiveness of such a scheme from the point of view of the government.

The intangibility of microinsurance makes it difficult to attract potential consumers. Most policyholders will never experience a claim and so cannot perceive the real value of microinsurance, paying more to the scheme (in terms of premium payments) than what they actually receive from it. It is only when claims are settled that microinsurance becomes tangible. The alternative microinsurance subsidy scheme described here could increase client value, since, for example, individuals below the barrier $B$ may submit claims, receive a payout and therefore perceive the value of microinsurance when they suffer a loss, regardless of whether they have ever paid a single premium. Further ways of increasing microinsurance client value include bundling microinsurance with other products and introducing Value Added Services (VAS), which (for health schemes) are services such as telephone hotlines for consultation with doctors or remote diagnosis services, offered to clients outside of the microinsurance contract \citep{Article:Madhur2019}. 

\vspace{0.3cm}

\begin{proposition}\label{GeneralSetting-Subsection61-Proposition1}

Consider a household enrolled in an alternative microinsurance scheme with subsidised flexible premiums, capital barrier $B \ge x^{\scaleto{(\mathcal{A})*}{5pt}}$ and proportionality factor $1-\kappa \in [0,1]$. Assume an initial capital $x \ge x^{\scaleto{(\mathcal{A})*}{5pt}}$, capital growth rates $r^{\scaleto{(\kappa)}{5pt}}$ and $r$ above and below the barrier, respectively, intensity $\lambda > 0$ and exponentially distributed capital losses with parameter $\alpha^{\scaleto{(\kappa)}{5pt}} > 0$. The Laplace transform of the trapping time is given by

\vspace{0.3cm}

\footnotesize
\begin{align}
m_{\delta}^{\scaleto{(\mathcal{A})}{5pt}}(x) =
\begin{cases}                 C_{1}M\left(-\frac{\delta}{r},1-\frac{\lambda+\delta}{r};y^{\scaleto{(\mathcal{A})}{5pt}}(x)\right) + C_{2}e^{y^{\scaleto{(\mathcal{A})}{5pt}}(x)}U\left(1-\frac{\lambda}{r}, 1-\frac{\lambda+\delta}{r};-y^{\scaleto{(\mathcal{A})}{5pt}}(x)\right) & \text{for $ x^{\scaleto{(\mathcal{A})*}{5pt}} \leq x \leq B$}, \\                C_{3}M\left(-\frac{\delta}{r^{\scaleto{(\kappa)}{5pt}}},1-\frac{\lambda+\delta}{r^{\scaleto{(\kappa)}{5pt}}};y^{\scaleto{(\mathcal{A})}{5pt}}(x)\right)
+ C_{4}e^{y^{\scaleto{(\mathcal{A})}{5pt}}(x)}U\left(1-\frac{\lambda}{r^{\scaleto{(\kappa)}{5pt}}}, 1-\frac{\lambda+\delta}{r^{\scaleto{(\kappa)}{5pt}}};-y^{\scaleto{(\mathcal{A})}{5pt}}(x)\right) & \text{for  $x \geq B$},
\end{cases}
\\\label{GeneralSetting-Subsection61-Equation1}
\end{align}
\normalsize

\vspace{0.3cm}

where $y^{\scaleto{(\mathcal{A})}{5pt}}(x)=-\alpha^{\scaleto{(\kappa)}{5pt}} (x-x^{\scaleto{(\mathcal{A})*}{5pt}})$ and the constants $C_{i}$ for $i=1,2,3,4$ are given by \eqref{Appendix A: Mathematical Proofs-Equation22}, \eqref{Appendix A: Mathematical Proofs-Equation26}, \eqref{Appendix A: Mathematical Proofs-Equation21} and \eqref{Appendix A: Mathematical Proofs-Equation25}, respectively.
\end{proposition}

\vspace{0.3cm}

A detailed mathematical proof of Proposition \ref{GeneralSetting-Subsection61-Proposition1} is provided in Appendix \ref{ProofofProposition6.1}.

\vspace{0.3cm}

    \begin{remark}
    
    The trapping probability $\psi^{\scaleto{(\mathcal{A})}{5pt}}(x)$ for the alternative microinsurance subsidy scheme is given by
    
    \vspace{0.3cm}
    
    \begin{align}
    \psi^{\scaleto{(\mathcal{A})}{5pt}}(x)=\begin{cases}  1 - \frac{\Gamma\left(\frac{\lambda}{r}\right)-\Gamma\left(\frac{\lambda}{r};-y^{\scaleto{(\mathcal{A})}{5pt}}(x)\right)}{ (-y^{\scaleto{(\mathcal{A})}{5pt}}(B))^{\lambda \left(\frac{1}{r}-\frac{1}{r^{\scaleto{(\kappa)}{3.5pt}}}\right)}\Gamma \left(\frac{\lambda}{r^{\scaleto{(\kappa)}{3.5pt}}}; - y^{\scaleto{(\mathcal{A})}{5pt}}(B)\right) +\Gamma\left(\frac{\lambda}{r}\right)-\Gamma \left(\frac{\lambda}{r}; - y^{\scaleto{(\mathcal{A})}{5pt}}(B)\right)}  & \textit{for $ x^{\scaleto{(\mathcal{A})*}{5pt}} \leq x \leq B$}, \\
    \frac{(-y^{\scaleto{(\mathcal{A})}{5pt}}(B))^{\lambda \left(\frac{1}{r}-\frac{1}{r^{\scaleto{(\kappa)}{3.5pt}}}\right)} \Gamma \left(\frac{\lambda}{r^{\scaleto{(\kappa)}{3.5pt}}}; - y^{\scaleto{(\mathcal{A})}{5pt}}(x)\right)}{ (-y^{\scaleto{(\mathcal{A})}{5pt}}(B))^{\lambda \left(\frac{1}{r}-\frac{1}{r^{\scaleto{(\kappa)}{3.5pt}}}\right)}\Gamma \left(\frac{\lambda}{r^{\scaleto{(\kappa)}{3.5pt}}}; - y^{\scaleto{(\mathcal{A})}{5pt}}(B)\right) +\Gamma\left(\frac{\lambda}{r}\right)-\Gamma \left(\frac{\lambda}{r}; - y^{\scaleto{(\mathcal{A})}{5pt}}(B)\right)} & \textit{for  $x \geq B$}.
    \end{cases}
    \label{GeneralSetting-Subsection61-Equation2}
\end{align}

\vspace{0.3cm}

Similar to the subsidised case, the optimal barrier $B$ can be found by determining the solution of the equation

\vspace{0.3cm}

\begin{align}
    \psi^{\scaleto{(\mathcal{A})}{5pt}}(x)=\psi(x),   \label{GeneralSetting-Subsection61-Equation3}
\end{align}

\vspace{0.3cm}

where $\psi^{\scaleto{(\mathcal{A})}{5pt}}(x)$ and $\psi(x)$ denote the trapping probabilities of the capital process under the alternative microinsurance subsidy scheme and the uninsured capital process, respectively. Some examples for varying initial capital are presented at the end of this section. 

\end{remark} 

\vspace{0.3cm}

\begin{remark}

When $B \rightarrow x^{\scaleto{(\mathcal{A})*}{5pt}}$, the trapping probability for the alternative microinsurance subsidy scheme is equal to the trapping probability obtained for the insured case $\psi^{\scaleto{(\kappa)}{5pt}}(x)$:

\vspace{0.3cm}

\begin{align}
    \lim_{B \rightarrow x^{\scaleto{(\mathcal{A})*}{3.5pt}}}\psi^{\scaleto{(\mathcal{A})}{5pt}}(x)=\frac{\Gamma\left(\frac{\lambda}{r^{\scaleto{(\kappa)}{5pt}}}; -y^{\scaleto{(\kappa)}{5pt}}(x)\right)}{\Gamma\left(\frac{\lambda}{r^{\scaleto{(\kappa)}{5pt}}}\right)}.
    \label{GeneralSetting-Subsection61-Equation4}
\end{align}

\vspace{0.3cm}

Moreover, when $B \rightarrow \infty$, the trapping probability is given by

\vspace{0.3cm}

\begin{align}
    \lim_{B \rightarrow \infty}\psi^{\scaleto{(\mathcal{A})}{5pt}}(x)=\frac{\Gamma\left(\frac{\lambda}{r}; -y^{\scaleto{(\kappa)}{5pt}}(x)\right)}{\Gamma\left(\frac{\lambda}{r}\right)},
    \label{GeneralSetting-Subsection61-Equation5}
\end{align}

\vspace{0.3cm}

which is exactly the trapping probability of the insured subsidised process $\psi^{\scaleto{\pi^{*}(\kappa,\theta)}{5pt}}(x)$ with $\pi^{*}=0$. 

\end{remark} 

\vspace{0.3cm}

\begin{remark}
Figure \ref{GeneralSetting-Subsection61-Figure1} displays the expected trapping time under the alternative microinsurance subsidy scheme for varying initial capital. Again, in line with intuition, the expected trapping time is an increasing function of both the capital level $B$ and initial capital $x$. Steps for obtaining the expected trapping time under the alternative microinsurance subsidy scheme are very similar to those used to derive equation \eqref{TheTrappingTime-Section3-Equation9}.

\begin{figure}[H]
    \centering
	\includegraphics[width=8cm, height=8cm]{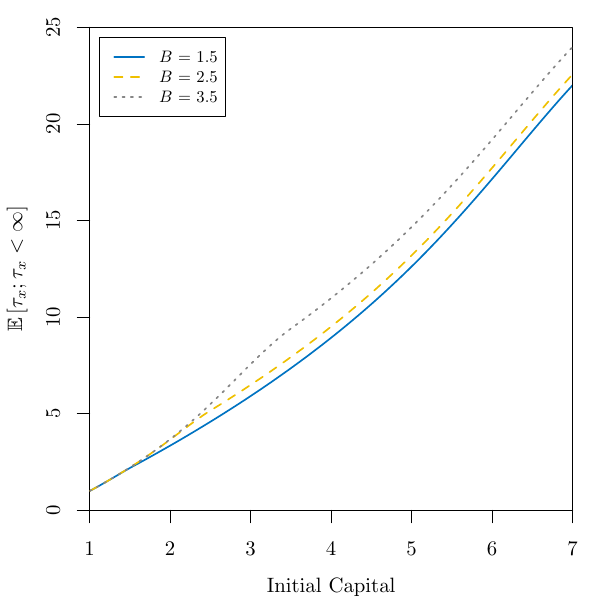}
	\caption{Expected trapping time when $Z_{i} \sim Exp(1)$, $a = 0.8$, $b = 1.4$, $c = 0.4$, $\lambda = 1$, $x^{\scaleto{(\mathcal{A})*}{5pt}} = 1$, $\kappa = 0.5$, and $\theta=0.5$ for $B=1.5,2.5,3.5$.}
	\label{GeneralSetting-Subsection61-Figure1}
\end{figure}

\end{remark}

\begin{figure}[H]
	\begin{subfigure}[b]{0.5\linewidth}
  		\includegraphics[width=8cm, height=8cm]{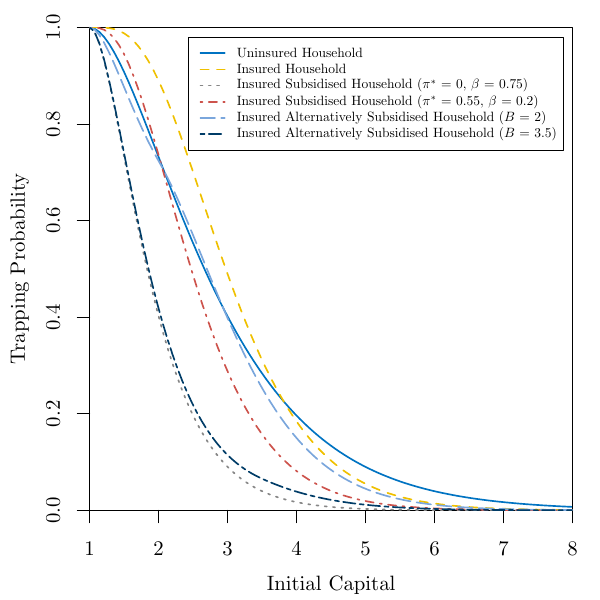}
		\caption{}
  		\label{GeneralSetting-Subsection61-Figure2-a}
	\end{subfigure}
	\begin{subfigure}[b]{0.5\linewidth}
  		\includegraphics[width=8cm, height=8cm]{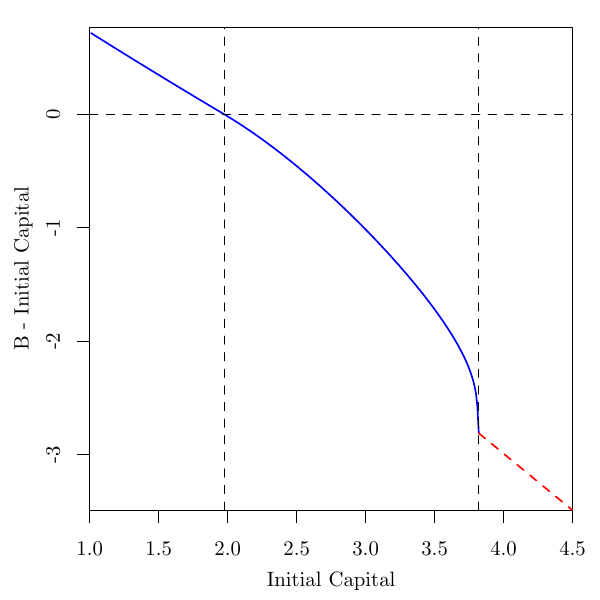}
		\caption{}
  		\label{GeneralSetting-Subsection61-Figure2-b}
	\end{subfigure}
	\caption{(a) Trapping probabilities for the uninsured, insured, insured subsidised with $\pi^{*}=0, 0.55$ and insured alternatively subsidised with $B = 2, 3.5$ capital processes when $Z_{i} \sim Exp(1)$, $a = 0.1$, $b = 1.4$, $c = 0.4$, $\lambda = 1$, $x^{*} = x^{\scaleto{(\kappa)*}{5pt}} = x^{\scaleto{\pi(\kappa, \theta)*}{5pt}} = x^{\scaleto{(\mathcal{A})*}{5pt}} = 1$, $\kappa = 0.5$, $\theta = 0.5$ and $\pi = 0.75$ (b) Difference between the optimal barrier and the initial capital, i.e. $B-x$, for varying initial capital, when $Z_{i} \sim Exp(1)$, $a = 0.1$, $b = 1.4$, $c = 0.4$, $\lambda = 1$, $x^{\scaleto{(\mathcal{A})*}{5pt}} = 1$, $\kappa = 0.5$, $\theta = 0.5$ and $\pi = 0.75$.}
	\label{GeneralSetting-Subsection61-Figure2}
\end{figure}

Figure \ref{GeneralSetting-Subsection61-Figure2-a} presents the trapping probabilities for varying initial capital under the uninsured, insured, insured subsidised and insured alternatively subsidised schemes. As expected, increasing the value of the capital barrier $B$ helps households to reduce their probability of falling into the area of poverty, since support from the government is received when their capital resides in the region between the critical capital $x^{\scaleto{(\mathcal{A})*}{5pt}}$ and the capital level $B$. Furthermore, as in the previous section, households with higher levels of initial capital do not need government support, insurance without subsidies decreases their trapping probability to a level below the uninsured (households with initial capital greater than or equal to the point at which the yellow short-dashed line intersects the blue solid line). The optimal barrier for these households is in fact the critical capital, i.e. $B=x^{\scaleto{(\mathcal{A})*}{5pt}}$, this household group can therefore afford to cover the costs of microinsurance coverage themselves.

Figure \ref{GeneralSetting-Subsection61-Figure2-b} shows that for the most vulnerable, governments should set the barrier level $B$ above their initial capital to remove capital growth constraints associated with premium payments. This level should be selected until the household reaches a capital level that is adequate in ensuring their trapping probability is equal to that of an uninsured household. Conversely, for more privileged households (middle area of Figure \ref{GeneralSetting-Subsection61-Figure2-b}), the government should establish barriers below their initial capital, with households paying premiums as soon as they enrol in the microinsurance scheme. This behaviour of the optimal barrier is mainly due to the fact that the capital level of such households is distant from the critical capital $x^{\scaleto{(\mathcal{A})*}{5pt}}$. These households are unlikely to fall into the area of poverty after suffering one (non-catastrophe) capital loss, they are instead likely to fall into the region between the critical capital $x^{\scaleto{(\mathcal{A})*}{5pt}}$ and the barrier level $B$ (the area within which the government pays microinsurance premiums), before entering the area of poverty. Thus, the aforementioned region acts as a  \lq \lq buffer\rq \rq, with households in this region benefiting from coverage without the need for premium payments. Increasing initial capital leads to a decrease in the size of the \lq \lq buffer\rq \rq \ region until it disappears when the optimal barrier $B=x^{\scaleto{(\mathcal{A})*}{5pt}}$, as shown by the red dashed line in the right area of Figure \ref{GeneralSetting-Subsection61-Figure2-b}.

\subsection{Cost of Social Protection}\label{CostofSocialProtection-Subsection62}

Similar to the previous section, it is reasonable to measure the governmental cost-effectiveness of providing microinsurance premium subsidies to households under the alternative microinsurance subsidy scheme. For this reason, we define $\tau^{\scaleto{(\mathcal{A})}{5pt}}_{x}$ as the trapping time of a household covered by the alternative subsidised microinsurance scheme and $V^{\scaleto{(\mathcal{A})}{5pt}}(x)$ as the expectation of the present value of all subsidies provided by the government to the household until the trapping time, that is

\vspace{0.3cm}

\begin{equation}
V^{\scaleto{(\mathcal{A})}{5pt}}(x):=\mathbb{E}\left[\int_{0}^{\tau^{\scaleto{(\mathcal{A})}{3pt}}_{x}} \pi e^{-\delta t} \mathbbm{1}_{\left\{X_{t}^{\scaleto{(\mathcal{A})}{5pt}} < B\right\}} dt \middle|  X^{\scaleto{(\mathcal{A})}{5pt}}_{0}=x\right].
\label{CostofSocialProtection-Subsection62-Equation1}
\end{equation}

\vspace{0.3cm}

In this article, estimates for $V^{\scaleto{(\mathcal{A})}{5pt}}(x)$ are produced via Monte Carlo simulation.

\vspace{0.3cm}

\begin{remark}
As in Section \ref{CostofSocialProtection-Subsection52}, we can easily derive an expression for $m_{\delta,w}^{\scaleto{(\mathcal{A})}{5pt}}(x)$, the expected discounted cost incurred by the government at the trapping time under the alternative microinsurance scheme.
\end{remark}

Following a similar procedure to that in the proof of Proposition \ref{CostofSocialProtection-Subsection52-Proposition2} (details of which are shown in Appendix \ref{Appendix A: Mathematical Proofs}), but for the alternative microinsurance subsidy scheme, one obtains the expected discounted cost incurred by the government at the trapping time.

\vspace{0.3cm}

\begin{proposition}\label{CostofSocialProtection-Subsection62-Proposition2}

Consider a household enrolled in an alternative microinsurance scheme with subsidised flexible premiums, capital barrier $B \ge x^{\scaleto{(\mathcal{A})*}{5pt}}$ and proportionality factor $1-\kappa \in [0,1]$. Assume an initial capital $x \ge x^{\scaleto{(\mathcal{A})*}{5pt}}$, capital growth rates $r^{\scaleto{(\kappa)}{5pt}}$ and $r$ above and below the barrier, respectively, intensity $\lambda > 0$, exponentially distributed capital losses with parameter $\alpha^{\scaleto{(\kappa)}{5pt}} > 0$ and a cost to lift households further away from the area of poverty $M^{\scaleto{(\kappa)}{5pt}}-x^{\scaleto{(\mathcal{A})*}{5pt}}$, with $M^{\scaleto{(\kappa)}{5pt}} \geq x^{\scaleto{(\mathcal{A})*}{5pt}}$.  The expected discounted cost incurred by the government at the trapping time is

\vspace{0.3cm}

\begin{align}
m_{\delta,w}^{\scaleto{(\mathcal{A})}{5pt}}(x) = \left[\frac{1}{\alpha^{\scaleto{(\kappa)}{5pt}}}+M^{\scaleto{(\kappa)}{5pt}}-x^{\scaleto{(\mathcal{A})*}{5pt}}\right]m_{\delta}^{\scaleto{(\mathcal{A})}{5pt}}(x),
\label{CostofSocialProtection-Subsection62-Equation2}
\end{align}

\vspace{0.3cm}

where $m_{\delta}^{\scaleto{(\mathcal{A})}{5pt}}(x)$ is given by \eqref{GeneralSetting-Subsection61-Equation1}.
\end{proposition}

As for the subsidised scheme, under the alternative scheme, we consider the cost of social protection incurred by the government to be equal to the expected discounted subsidies provided until trapping plus the expected discounted cost incurred at trapping, here given by \eqref{CostofSocialProtection-Subsection62-Equation1} and \eqref{CostofSocialProtection-Subsection62-Equation2}, respectively.

\begin{figure}[H]
    \centering
	\includegraphics[width=8cm, height=8cm]{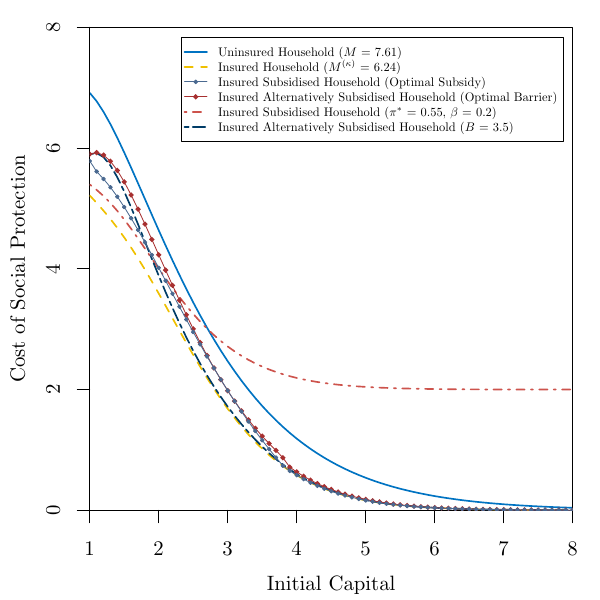}
	\caption{Cost of social protection for the uninsured, insured, insured subsidised with $\pi^{*}= \pi^{*}_{\scaleto{\text{ {\fontfamily{qcr}\selectfont
 Optimal}}}{4pt}}, 0.55$ and insured alternatively subsidised capital processes with $B=B_{\scaleto{\text{ {\fontfamily{qcr}\selectfont
 Optimal}}}{4pt}}, 3.5$ when $Z_{i}\sim Exp(1)$, $a = 0.1$, $b = 1.4$, $c = 0.4$, $\lambda=1$, $x^{*} = x^{\scaleto{(\kappa)*}{5pt}} = x^{\scaleto{\pi(\kappa, \theta)*}{5pt}} = x^{\scaleto{(\mathcal{A})*}{5pt}} = 1$, $\kappa = 0.5$, $\theta = 0.5$, $\delta =0.1$, $\epsilon = 0.01$ and $\pi= 0.75$.}
	\label{CostofSocialProtection-Subsection62-Figure1}
\end{figure}

Figure \ref{CostofSocialProtection-Subsection62-Figure1} compares the cost of social protection for the uninsured, insured, insured subsidised and insured alternatively subsidised households. Cost of social protection for the most vulnerable is reduced with all forms of microinsurance coverage (yellow dashed, red dashed-dotted, blue dashed-dashed, blue circular-marked and red diamond-marked lines are all below the blue solid line for initial capitals close to the critical capital $x^{*}$). This aligns with the high trapping probability associated with this portion of the population when uninsured, with governments almost surely needing to lift households out of the area of poverty. Although already eliminated when providing optimal subsidies under the insured subsidised scheme (blue circular-marked line below the red dashed-dotted line for the most privileged), the aforementioned drawback of governments subsidising premiums indefinitely almost surely under non-optimal subsidised schemes is also eliminated under both optimal and non-optimal alternative subsidy schemes due to the ceasing of subsidies on households reaching sufficient capital (red diamond-marked and blue dashed-dashed lines below red dashed-dotted line for households with higher levels of capital). Furthermore, as seen in Figure \ref{GeneralSetting-Subsection61-Figure2-a}, when the barrier level is sufficiently high all households observe a decrease in their trapping probability, almost reaching the trapping probability of a household enrolled in a fully subsidised insurance scheme with $\pi^{*}= 0$. However, even when high barrier levels are considered, under the alternatively subsidised scheme governments are not required to subsidise premiums indefinitely, since households will pay the entire premium once their capital reaches a sufficient level. This scheme thus makes it possible to reduce the trapping probability for any household, while reducing the cost of social protection incurred by the government, which highlights the cost-efficiency of this alternative scheme.

\section{Conclusion} \label{Conclusion-Section6}

Comparing the impact of three microinsurance frameworks on the trapping probabilities of low-income households, we provide evidence for the importance of governmentally supported inclusive insurance in the strive towards poverty alleviation. The results of Sections \ref{IntroducingMicroinsurance-Section4} and \ref{MicroinsurancewithSubsidisedConstantPremiums-Section5} support those of \cite{Article:Kovacevic2011}, highlighting a threshold below which insurance could increase the probability of trapping. Motivated by the recent increased involvement of governments in the support of insurance programmes and maintaining the idea of \ \lq \lq smart\rq \rq \ subsidies, we have introduced a transparent method with a mathematical foundation for calculating \lq \lq optimal subsidies\rq \rq \ that can strengthen government social protection programmes while lowering the associated costs.

Numerical examples indicate that while the proposed insurance mechanisms (with or without subsidies) reduce the cost of social protection for the most vulnerable, they do not reduce their probability of trapping. This undermines the faculty of inclusive insurance as a cost-effective social protection strategy for poverty alleviation and brings to light questions as to its capability in reducing both the probability of households falling below the poverty line and the associated social protection costs. However, our analysis of a subsidised microinsurance scheme with a premium payment barrier suggests that in general, the trapping probability of a household covered by such a scheme is reduced in comparison to when covered by unsubsidised and (for the most vulnerable) partially subsidised microinsurance schemes, in addition to when uninsured, alleviating this limitation.

More significant influence is observed in relation to the governmental cost of social protection, with the cease of subsidy payments when household capital is sufficient facilitating government savings and therefore increasing social protection efficiency, thus evidencing the relevance of the alternative scheme as a cost-effective social protection strategy for poverty reduction. The cost of social protection for those closest to the area of poverty remains lower than the corresponding uninsured cost in both subsidised frameworks considered, achieving similar results to those obtained with the targeted-subsidisy scheme proposed by \cite{Article:Janzen2020}. In our analysis, for such households, governments must account for their support of premium payments, the likely need for household removal from poverty and an extra capital injection to ensure they will not return to poverty with some level of confidence. Nevertheless, total subsidies paid by the government have a small weight within the cost of social protection due to the fact that those closest to the poverty line will fall into the area of poverty almost surely. The capital injection on trapping is also much lower in comparison to that of uninsured households. Each of these factors enhances the reduction in the cost of social protection for the most vulnerable.

Given the decrease in trapping probability and governmental cost of social protection under the barrier strategy scheme considered, this paper advocates for the development of public-private partnerships (PPPs) for the provision of affordable insurance. Through well-designed subsidy schemes PPPs can reduce vulnerability to poverty in a cost-effective manner. A key motivator for supporting the development of insurance mechanisms is their ability to improve productivity and access to resources. Insurance can improve access to new technologies, credit and hospital services, for example, smoothing the movement of low-income individuals along the economic cycle and providing them with the capacity to move out of poverty and to stay there. Insurance is therefore both productive and protective in preventing poverty. By making insurance more accessible, subsidies improve access to productive resources while providing protective cover.

In presenting the results of the analysis it is important to note the limitations of the adopted approach. All types of insurance are captured in our \lq \lq insurance\rq \rq \ coverage, we therefore do not consider the susceptibility of households to losses of varying severity which would align with the presence of different lines of business, i.e. health, life and agricultural insurance. In addition, the subsidy schemes are assumed to be continuous, with households receiving government subsidies forever in the constant case and while below the barrier in the flexible case. This raises questions in regard to the sustainability of such schemes. An alternative approach would be to phase out subsidisation over time, enabling households to experience and therefore understand the benefit of insurance such that they go on to purchase coverage once no longer subsidised. The barrier strategy would be difficult to implement in practice, requiring continuous tracking of a household’s capital over time. Future research will involve adjusting the assumption of random-valued losses to consider random-proportional losses, as in \cite{Article:Kovacevic2011}. Through the assumption of random-valued losses, a household\rq s level of accumulated capital could fall below zero. In this case, the household would lose more than what it has and would likely continue to lose capital even after surpassing the capital level of zero. Under the random-proportional losses assumption, the concept of trapping is better captured, as once within the area of poverty it is impossible to escape from either side. Moreover, this assumption fits the idea that the amount of capital lost by a household on experiencing an adverse event should depend on the amount of capital that the household currently possesses, e.g. when a household has little capital, then we would expect them to have less to lose.

The main takeaway from this analysis is the importance of the \lq \lq missing middle\rq \rq, i.e. those close to but above the poverty line, for whom paying the premium is a risk in itself. This takeaway is also applicable in traditional insurance markets. In comparison to large international enterprises, small and medium-sized enterprises (SMEs) with limited liquidity are more likely to require government support in the event of a severe loss. For example, in the pandemic context, the Covid-19 pandemic saw governments stepping in to cover the loss of jobs, wages and, in some cases, hospitalisation costs. Without stakeholder capital injections or such governmental support, SMEs would be extremely susceptible to insolvency. In addition, reinsurance premiums for such an extreme loss would be more severe for smaller enterprises, while larger enterprises are likely to be better posed to protect themselves against such severe risks. 

The \lq \lq missing middle\rq \rq \ exists in all fields of insurance, those closest to insolvency have less capacity to protect themselves against the occurrence of a catastrophic loss. Smart solutions supported by large organisations, including governments and intergovernmental organisations, should therefore be designed to mitigate the increased risk faced by the most vulnerable.

\section*{Funding} \label{Introduction-Section8}

The work of K.H and C.C was supported by Engineering and Physical Sciences Research Council (EPSRC) [grant number EP/W522399/1]; and the EPSRC and ESRC Centre for Doctoral Training on Quantification and Management of Risk and Uncertainty in Complex Systems Environments [grant number EP/L015927/1]

\setcitestyle{numbers} 

\bibliographystyle{chicago} 

\bibliography{main}

\begin{thebibliography}{}

\bibitem[\protect\citeauthoryear{Abramowitz and Stegun}{Abramowitz and
  Stegun}{1972}]{Book:Abramowitz1964}
Abramowitz, M. and I.~A. Stegun (1972).
\newblock {\em Handbook of Mathematical Functions with Formulas, Graphs, and
  Mathematical Tables}.
\newblock Washington, D.C.: U.S. Department of Commerce.

\bibitem[\protect\citeauthoryear{Albrecher, Beirlant, and Teugels}{Albrecher
  et~al.}{2017}]{Book:Albrecher2017}
Albrecher, H., J.~Beirlant, and J.~L. Teugels (2017).
\newblock {\em Reinsurance: Actuarial and Statistical Aspects}.
\newblock Oxford: John Wiley \& Sons.

\bibitem[\protect\citeauthoryear{Albrecher, Hartinger, and Tichy}{Albrecher
  et~al.}{2005}]{Article:Albrecher2005}
Albrecher, H., J.~Hartinger, and R.~F. Tichy (2005).
\newblock {On the Distribution of Dividend Payments and the Discounted Penalty
  Function in a Risk Model with Linear Dividend Barrier}.
\newblock {\em Scandinavian Actuarial Journal\/}~(2), 103--126.

\bibitem[\protect\citeauthoryear{Asmussen and Albrecher}{Asmussen and
  Albrecher}{2010}]{Book:Asmussen2010}
Asmussen, S. and H.~Albrecher (2010).
\newblock {\em Ruin Probabilities}.
\newblock Singapore: World Scientific.

\bibitem[\protect\citeauthoryear{Auzzir, Haigh, and Amaratunga}{Auzzir
  et~al.}{2014}]{Article:Auzzir2014}
Auzzir, Z.~A., R.~P. Haigh, and D.~Amaratunga (2014).
\newblock {Public-Private Partnerships (PPP) in Disaster Management in
  Developing Countries: A Conceptual Framework}.
\newblock {\em Procedia Economics and Finance\/}~{\em 18}, 807--814.

\bibitem[\protect\citeauthoryear{Aza{\"\i}s and Genadot}{Aza{\"\i}s and
  Genadot}{2015}]{Article:Azais2015}
Aza{\"\i}s, R. and A.~Genadot (2015).
\newblock {Semi-Parametric Inference for the Absorption Features of a
  Growth-Fragmentation Model}.
\newblock {\em TEST\/}~{\em 24\/}(2), 341--360.

\bibitem[\protect\citeauthoryear{Azariadis and Stachurski}{Azariadis and
  Stachurski}{2005}]{Inbook:Aghion2005}
Azariadis, C. and J.~Stachurski (2005).
\newblock {Poverty Traps}.
\newblock In P.~Aghion and S.~N. Durlauf (Eds.), {\em Handbook of Economic
  Growth}, Volume~22 of {\em Handbooks in Economics}, Chapter~5, pp.\
  295--384. North-Holland: Elsevier.

\bibitem[\protect\citeauthoryear{Bardhan and Udry}{Bardhan and
  Udry}{1999}]{Book:Bardhan1999}
Bardhan, P. and C.~Udry (1999).
\newblock {\em Development Microeconomics}.
\newblock Oxford: Oxford University Press.

\bibitem[\protect\citeauthoryear{Barrett, Garg, and McBride}{Barrett
  et~al.}{2016}]{Article:Barrett2016}
Barrett, C.~B., T.~Garg, and L.~McBride (2016).
\newblock {Well-Being Dynamics and Poverty Traps}.
\newblock {\em Annual Review of Resource Economics\/}~{\em 8\/}(1), 303--327.

\bibitem[\protect\citeauthoryear{Biener and Eling}{Biener and
  Eling}{2012}]{Article:Biener2012}
Biener, C. and M.~Eling (2012).
\newblock {Insurability in Microinsurance Markets: An Analysis of Problems and
  Potential Solutions}.
\newblock {\em The Geneva Papers on Risk and Insurance - Issues and
  Practice\/}~{\em 37\/}(1), 77--107.

\bibitem[\protect\citeauthoryear{Biener, Eling, and Schmit}{Biener
  et~al.}{2014}]{Article:Biener2014}
Biener, C., M.~Eling, and J.~T. Schmit (2014).
\newblock {Regulation in Microinsurance Markets: Principles, Practice, and
  Directions for Future Development}.
\newblock {\em World Development\/}~{\em 58}, 21--40.

\bibitem[\protect\citeauthoryear{Bowles, Durlauf, and Hoff}{Bowles
  et~al.}{2006}]{Book:Bowles2006}
Bowles, S., S.~N. Durlauf, and K.~Hoff (2006).
\newblock {\em Poverty Traps}.
\newblock New Jersey: Princeton University Press.

\bibitem[\protect\citeauthoryear{Carter and Janzen}{Carter and
  Janzen}{2018}]{Article:Carter2018}
Carter, M.~R. and S.~A. Janzen (2018).
\newblock {Social Protection in the Face of Climate Change: Targeting
  Principles and Financing Mechanisms}.
\newblock {\em Environment and Development Economics\/}~{\em 23\/}(3),
  369--389.

\bibitem[\protect\citeauthoryear{Chantarat, Mude, Barrett, and
  Turvey}{Chantarat et~al.}{2017}]{Article:Chantarat2017}
Chantarat, S., A.~G. Mude, C.~B. Barrett, and C.~G. Turvey (2017).
\newblock {Welfare Impacts of Index Insurance in the Presence of a Poverty
  Trap}.
\newblock {\em World Development\/}~{\em 94}, 119--138.

\bibitem[\protect\citeauthoryear{Churchill}{Churchill}{2007}]{Article:Churchill2007}
Churchill, C. (2007).
\newblock {Insuring the Low-Income Market: Challenges and Solutions for
  Commercial Insurers}.
\newblock {\em The Geneva Papers on Risk and Insurance - Issues and
  Practice\/}~{\em 32\/}(3), 401--412.

\bibitem[\protect\citeauthoryear{Churchill and Matul}{Churchill and
  Matul}{2012}]{Book:Churchill2012}
Churchill, C. and M.~Matul (2012).
\newblock {\em Protecting the Poor: A Microinsurance Compendium (Vol. 2)}.
\newblock Geneva: International Labour Organization (ILO).

\bibitem[\protect\citeauthoryear{Cole, Gin\'e, Tobacman, Topalova, Townsend,
  and Vickery}{Cole et~al.}{2013}]{Article:Cole2013}
Cole, S., X.~Gin\'e, J.~Tobacman, P.~Topalova, R.~Townsend, and J.~Vickery
  (2013).
\newblock {Barriers to Household Risk Management: Evidence From India}.
\newblock {\em American Economic Journal: Applied Economics\/}~{\em 5\/}(1),
  104--135.

\bibitem[\protect\citeauthoryear{Constantinescu, Ramirez, and
  Zhu}{Constantinescu et~al.}{2019}]{Article:Constantinescu2019}
Constantinescu, C.~D., J.~M. Ramirez, and W.~R. Zhu (2019).
\newblock {An Application of Fractional Differential Equations to Risk Theory}.
\newblock {\em Finance and Stochastics\/}~{\em 23\/}(4), 1001--1024.

\bibitem[\protect\citeauthoryear{Dasgupta}{Dasgupta}{1997}]{Article:Dasgupta1997}
Dasgupta, P. (1997).
\newblock {Nutritional Status, the Capacity for Work, and Poverty Traps}.
\newblock {\em Journal of Econometrics\/}~{\em 77\/}(1), 5--37.

\bibitem[\protect\citeauthoryear{Davis}{Davis}{1984}]{Article:Davis1984}
Davis, M. H.~A. (1984).
\newblock {Piecewise-Deterministic Markov Processes: A General Class of
  Non-Diffusion Stochastic Models}.
\newblock {\em Journal of the Royal Statistical Society: Series B
  (Methodological)\/}~{\em 46\/}(3), 353--388.

\bibitem[\protect\citeauthoryear{{De Weerdt} and Dercon}{{De Weerdt} and
  Dercon}{2006}]{Article:DeWeerdt2006}
{De Weerdt}, J. and S.~Dercon (2006).
\newblock {Risk-Sharing Networks and Insurance Against Illness}.
\newblock {\em Journal of Development Economics\/}~{\em 81\/}(2), 337--356.

\bibitem[\protect\citeauthoryear{Dercon, {De Weerdt}, Bold, and
  Pankhurst}{Dercon et~al.}{2006}]{Article:Dercon2006}
Dercon, S., J.~{De Weerdt}, T.~Bold, and A.~Pankhurst (2006).
\newblock {Group-Based Funeral Insurance in Ethiopia and Tanzania}.
\newblock {\em World Development\/}~{\em 34\/}(4), 685--703.

\bibitem[\protect\citeauthoryear{Dror}{Dror}{2019}]{Article:Dror2019}
Dror, D.~M. (2019).
\newblock {Microinsurance: A Short History}.
\newblock {\em International Social Security Review\/}~{\em 72\/}(4), 107--126.

\bibitem[\protect\citeauthoryear{Eling, Pradhan, and Schmit}{Eling
  et~al.}{2014}]{Article:Eling2014}
Eling, M., S.~Pradhan, and J.~T. Schmit (2014).
\newblock {The Determinants of Microinsurance Demand}.
\newblock {\em The Geneva Papers on Risk and Insurance - Issues and
  Practice\/}~{\em 39\/}(2), 224--263.

\bibitem[\protect\citeauthoryear{Gartner, Shaver, Carter, and Reynolds}{Gartner
  et~al.}{2004}]{Book:Gartner2004}
Gartner, W.~B., K.~G. Shaver, N.~M. Carter, and P.~D. Reynolds (2004).
\newblock {\em Handbook of Entrepreneurial Dynamics: The Process of Business
  Creation}.
\newblock California: SAGE Publications, Inc.

\bibitem[\protect\citeauthoryear{Gerber and Shiu}{Gerber and
  Shiu}{1998}]{Article:Gerber1998}
Gerber, H.~U. and E.~S.~W. Shiu (1998).
\newblock {On the Time Value of Ruin}.
\newblock {\em North American Actuarial Journal\/}~{\em 2\/}(1), 48--72.

\bibitem[\protect\citeauthoryear{Harrison}{Harrison}{1977}]{Article:Harrison1977}
Harrison, J.~M. (1977).
\newblock {Ruin Problems with Compounding Assets}.
\newblock {\em Stochastic Processes and their Applications\/}~{\em 5\/}(1),
  67--79.

\bibitem[\protect\citeauthoryear{Hazell and Varangis}{Hazell and
  Varangis}{2020}]{Article:Hazell2020}
Hazell, P. and P.~Varangis (2020).
\newblock {Best Practices for Subsidizing Agricultural Insurance}.
\newblock {\em Global Food Security\/}~{\em 25\/}(1), 100326.

\bibitem[\protect\citeauthoryear{Hill, Gajate-Garrido, Phily, and Aparna}{Hill
  et~al.}{2014}]{Book:Hill2014}
Hill, R.~V., G.~Gajate-Garrido, C.~Phily, and D.~Aparna (2014).
\newblock {\em {Using Subsidies for Inclusive Insurance: Lessons From
  Agriculture and Health}}.
\newblock Microinsurance Paper No. 29. Geneva, Switzerland: International
  Labour Organization (ILO).

\bibitem[\protect\citeauthoryear{Ikegami, Carter, Barrett, and Janzen}{Ikegami
  et~al.}{2018}]{Inbook:Barrett2016}
Ikegami, M., M.~R. Carter, C.~B. Barrett, and S.~A. Janzen (2018).
\newblock {Poverty Traps and the Social Protection Paradox}.
\newblock In C.~B. Barrett, M.~R. Carter, and J.~P. Chavas (Eds.), {\em The
  Economics of Poverty Traps}, Chapter~6, pp.\  223--256. Chicago: University
  of Chicago Press.

\bibitem[\protect\citeauthoryear{Janzen, Carter, and Ikegami}{Janzen
  et~al.}{2021}]{Article:Janzen2020}
Janzen, S.~A., M.~R. Carter, and M.~Ikegami (2021).
\newblock {Can Insurance Alter Poverty Dynamics and Reduce the Cost of Social
  Protection in Developing Countries?}
\newblock {\em Journal of Risk and Insurance\/}~{\em 88\/}(2), 293--324.

\bibitem[\protect\citeauthoryear{Jensen, Ikegami, and Mude}{Jensen
  et~al.}{2017}]{Article:Jensen2017}
Jensen, N., M.~Ikegami, and A.~Mude (2017).
\newblock {Integrating Social Protection Strategies for Improved Impact: A
  Comparative Evaluation of Cash Transfers and Index Insurance in Kenya}.
\newblock {\em The Geneva Papers on Risk and Insurance - Issues and
  Practice\/}~{\em 42\/}(4), 675--707.

\bibitem[\protect\citeauthoryear{Kaur, Raj, Singh, and Chattu}{Kaur
  et~al.}{2021}]{Article:Kaur2021}
Kaur, S., H.~Raj, H.~Singh, and V.~K. Chattu (2021).
\newblock {Crop Insurance Policies in India: An Empirical Analysis of Pradhan
  Mantri Fasal Bima Yojana}.
\newblock {\em Risks\/}~{\em 9\/}(11).

\bibitem[\protect\citeauthoryear{Kousky, Wiley, and Shabman}{Kousky
  et~al.}{2021}]{Article:Kousky2021}
Kousky, C., H.~Wiley, and L.~Shabman (2021).
\newblock {Can Parametric Microinsurance Improve the Financial Resilience of
  Low-Income Households in the United States?}
\newblock {\em Economics of Disasters and Climate Change\/}~{\em 5\/}(3),
  301--327.

\bibitem[\protect\citeauthoryear{Kovacevic and Pflug}{Kovacevic and
  Pflug}{2011}]{Article:Kovacevic2011}
Kovacevic, R.~M. and G.~C. Pflug (2011).
\newblock {Does Insurance Help to Escape the Poverty Trap? — A Ruin Theoretic
  Approach}.
\newblock {\em Journal of Risk and Insurance\/}~{\em 78\/}(4), 1003--1027.

\bibitem[\protect\citeauthoryear{Kovacevic and Semmler}{Kovacevic and
  Semmler}{2021}]{Inbook:Kovacevic2021}
Kovacevic, R.~M. and W.~Semmler (2021).
\newblock {Poverty Traps and Disaster Insurance in a Bi-level Decision
  Framework}.
\newblock In J.~L. Haunschmied, R.~M. Kovacevic, W.~Semmler, and V.~M. Veliov
  (Eds.), {\em Dynamic Economic Problems with Regime Switches}, Volume~25 of
  {\em Dynamic Modeling and Econometrics in Economics and Finance}, Chapter~3,
  pp.\  57--83. Switzerland: Springer Nature.

\bibitem[\protect\citeauthoryear{Kraay and McKenzie}{Kraay and
  McKenzie}{2014}]{Article:Kraay2014}
Kraay, A. and D.~McKenzie (2014).
\newblock {Do Poverty Traps Exist? Assessing the Evidence}.
\newblock {\em Journal of Economic Perspectives\/}~{\em 28\/}(3), 127--148.

\bibitem[\protect\citeauthoryear{Kramer, Hazell, Alderman, Ceballos, Kumar, and
  Timu}{Kramer et~al.}{2022}]{Article:Kramer2022}
Kramer, B., P.~Hazell, H.~Alderman, F.~Ceballos, N.~Kumar, and A.~G. Timu
  (2022).
\newblock {Is Agricultural Insurance Fulfilling Its Promise for the Developing
  World? A Review of Recent Evidence}.
\newblock {\em Annual Review of Resource Economics\/}~{\em 14\/}(1), 291--311.

\bibitem[\protect\citeauthoryear{Kummer}{Kummer}{1837}]{Article:Kummer1837}
Kummer, E.~E. (1837).
\newblock {De integralibus quibusdam definitis et seriebus infinitis}.
\newblock {\em Journal f{\"u}r die reine und angewandte Mathematik\/}~(17),
  228--242.

\bibitem[\protect\citeauthoryear{Kyprianou}{Kyprianou}{2013}]{Book:Kyprianou2013}
Kyprianou, A.~E. (2013).
\newblock {\em Gerber–Shiu Risk Theory}.
\newblock Switzerland: Springer International Publishing.

\bibitem[\protect\citeauthoryear{Liao, Zhou, and Fan}{Liao
  et~al.}{2020}]{Article:Liao2020}
Liao, P., X.~Zhou, and Q.~Fan (2020).
\newblock {Does Agricultural Insurance Help Farmers Escape the Poverty Trap?
  Research Based on Multiple Equilibrium Models}.
\newblock {\em The Geneva Papers on Risk and Insurance - Issues and
  Practice\/}~{\em 45\/}(1), 203--223.

\bibitem[\protect\citeauthoryear{Linnerooth-Bayer and Mechler}{Linnerooth-Bayer
  and Mechler}{2007}]{Article:Linnerooth-Bayer2007}
Linnerooth-Bayer, J. and R.~Mechler (2007).
\newblock {Disaster Safety Nets for Developing Countries: Extending
  Public—Private Partnerships}.
\newblock {\em Environmental Hazards\/}~{\em 7\/}(1), 54--61.

\bibitem[\protect\citeauthoryear{Liu and Myers}{Liu and
  Myers}{2016}]{Article:Liu2016}
Liu, Y. and R.~J. Myers (2016).
\newblock {The Dynamics of Microinsurance Demand in Developing Countries Under
  Liquidity Constraints and Insurer Default Risk}.
\newblock {\em Journal of Risk and Insurance\/}~{\em 83\/}(1), 121--138.

\bibitem[\protect\citeauthoryear{Madhur and Saha}{Madhur and
  Saha}{2019}]{Article:Madhur2019}
Madhur, S.~K. and S.~Saha (2019).
\newblock {Protecting the Economic Health of the Poor in India: Are Health
  Mutuals the Right Medicine?}
\newblock {\em Development Policy Review\/}~{\em 37\/}(6), 843--853.

\bibitem[\protect\citeauthoryear{Matsuyama}{Matsuyama}{2008}]{Inbook:Matsuyama2008}
Matsuyama, K. (2008).
\newblock {Poverty Traps}.
\newblock In S.~N. Durlauf and L.~E. Blume (Eds.), {\em The New Palgrave
  Dictionary of Economics}. London: Palgrave Macmillan.

\bibitem[\protect\citeauthoryear{Paulsen}{Paulsen}{1998}]{Article:Paulsen1998}
Paulsen, J. (1998).
\newblock {Ruin Theory with Compounding Assets — A Survey}.
\newblock {\em Insurance: Mathematics and Economics\/}~{\em 22\/}(1), 3--16.

\bibitem[\protect\citeauthoryear{Platteau, {De Bock}, and Gelade}{Platteau
  et~al.}{2017}]{Article:Platteau2017}
Platteau, J.~P., O.~{De Bock}, and W.~Gelade (2017).
\newblock {The Demand for Microinsurance: A Literature Review}.
\newblock {\em World Development\/}~{\em 94}, 139--156.

\bibitem[\protect\citeauthoryear{Ramsay and Arcila}{Ramsay and
  Arcila}{2013}]{Article:Ramsay2013}
Ramsay, C.~M. and L.~D. Arcila (2013).
\newblock {Pricing Funeral (Burial) Insurance in a Microinsurance World with
  Emphasis on Africa}.
\newblock {\em North American Actuarial Journal\/}~{\em 17\/}(1), 63--81.

\bibitem[\protect\citeauthoryear{Sattayatham, Sangaroon, and
  Klongdee}{Sattayatham et~al.}{2013}]{Article:Sattayatham2013}
Sattayatham, P., K.~Sangaroon, and W.~Klongdee (2013).
\newblock {Ruin Probability-Based Initial Capital of the Discrete-Time Surplus
  Process}.
\newblock {\em Variance: Advancing the Science of Risk: CAS\/}~{\em 7\/}(1),
  74--81.

\bibitem[\protect\citeauthoryear{Seaborn}{Seaborn}{1991}]{Book:Seaborn1991}
Seaborn, J.~B. (1991).
\newblock {\em Hypergeometric Functions and Their Applications}.
\newblock New York: Springer-Verlag.

\bibitem[\protect\citeauthoryear{Segerdahl}{Segerdahl}{1942}]{Article:Segerdahl1942}
Segerdahl, C.~O. (1942).
\newblock {\"U}ber einige risikotheoretische fragestellungen.
\newblock {\em Scandinavian Actuarial Journal\/}~(1-2), 43--83.

\bibitem[\protect\citeauthoryear{Slater}{Slater}{1960}]{Book:Slater1960}
Slater, L.~J. (1960).
\newblock {\em Confluent Hypergeometric Functions}.
\newblock New York: Cambridge University Press.

\bibitem[\protect\citeauthoryear{Sundt and Teugels}{Sundt and
  Teugels}{1995}]{Article:Sundt1995}
Sundt, B. and J.~L. Teugels (1995).
\newblock {Ruin Estimates Under Interest Force}.
\newblock {\em Insurance: Mathematics and Economics\/}~{\em 16\/}(1), 7--22.

\bibitem[\protect\citeauthoryear{Townsend}{Townsend}{1994}]{Article:Townsend1994}
Townsend, R.~M. (1994).
\newblock {Risk and Insurance in Village India}.
\newblock {\em Econometrica\/}~{\em 62\/}(3), 539--591.

\bibitem[\protect\citeauthoryear{Tricomi}{Tricomi}{1947}]{Article:Tricomi1947}
Tricomi, F. (1947).
\newblock {Sulle funzioni ipergeometriche confluenti}.
\newblock {\em Annali di Matematica Pura ed Applicata\/}~{\em 26\/}(1),
  141--175.

\bibitem[\protect\citeauthoryear{Wang, Shi, Ye, Liu, and Zhou}{Wang
  et~al.}{2011}]{Article:Wang2011}
Wang, M., P.~Shi, T.~Ye, M.~Liu, and M.~Zhou (2011).
\newblock {Agriculture Insurance in China: History, Experience, and Lessons
  Learned}.
\newblock {\em International Journal of Disaster Risk Science\/}~{\em 2\/}(2),
  10--22.

\bibitem[\protect\citeauthoryear{Will, Groeneveld, Frank, and Müller}{Will
  et~al.}{2021}]{Article:Will2021}
Will, M., J.~Groeneveld, K.~Frank, and B.~Müller (2021).
\newblock {Informal Risk-Sharing Between Smallholders May Be Threatened By
  Formal Insurance: Lessons From a Stylized Agent-Based Model}.
\newblock {\em PLOS ONE\/}~{\em 16\/}(3), 1--18.

\bibitem[\protect\citeauthoryear{Ye, Hu, Barnett, Wang, and Gao}{Ye
  et~al.}{2020}]{Article:Ye2020}
Ye, T., W.~Hu, B.~J. Barnett, J.~Wang, and Y.~Gao (2020).
\newblock {Area Yield Index Insurance or Farm Yield Crop Insurance? Chinese
  Perspectives on Farmers' Welfare and Government Subsidy Effectiveness}.
\newblock {\em Journal of Agricultural Economics\/}~{\em 71\/}(1), 144--164.

\end{thebibliography}

\setcitestyle{authoryear}


\section*{Appendices}
\addcontentsline{toc}{section}{Appendices}
\renewcommand{\thesubsection}{\Alph{subsection}}
\setcounter{subsection}{0}

\numberwithin{equation}{subsubsection}

\subsection{Mathematical Proofs}\label{Appendix A: Mathematical Proofs}

\subsubsection{Proof of Proposition \ref{TheTrappingTime-Section3-Proposition1}} \label{ProofofProposition3.1}

Using standard arguments based on the infinitesimal generator, the expected discounted penalty function at the trapping time $m_{\delta}(x)$ as defined in \eqref{TheTrappingTime-Section3-Equation2}, can be characterised as the solution of the IDE

\vspace{0.3cm}

\begin{equation}
    r(x-x^{*})m_{\delta}'(x)-(\lambda + \delta) m_{\delta}(x)+\lambda \int_{0}^{x-x^{*}}m_{\delta}(x-z)dG_{Z}(z)=-\lambda A(x), \qquad x \ge x^{*},
    \label{Appendix A: Mathematical Proofs-Equation1}
\end{equation}

\vspace{0.3cm}

where

\vspace{0.3cm}

\begin{align}
    A(x) := \int_{x-x^{*}}^{\infty}w(x-x^{*}, z-(x-x^{*}))dG_{Z}(z).
    \label{Appendix A: Mathematical Proofs-Equation2}
\end{align}

\vspace{0.3cm}

When $Z_{i}\sim Exp(\alpha)$ and $w(x_{1}, x_{2})=1$, equation \eqref{Appendix A: Mathematical Proofs-Equation1} can be written such that

\vspace{0.3cm}

\begin{align}
    r(x-x^{*})m_{\delta}'(x)-(\lambda + \delta) m_{\delta}(x)+\lambda \int_{0}^{x-x^{*}}m_{\delta}(x-z)\alpha e^{-\alpha z} dz&=-\lambda e^{-\alpha (x-x^{*})} , \hspace{0.2cm} x \ge x^{*}. \\
    \label{Appendix A: Mathematical Proofs-Equation3}
\end{align}

\vspace{0.3cm}

Applying the operator $\left(\frac{d}{dx}+\alpha\right)$ to both sides of \eqref{Appendix A: Mathematical Proofs-Equation3}, together with a number of algebraic manipulations, yields the second order homogeneous differential equation

\vspace{0.3cm}

\begin{align}
    -\frac{(x-x^{*})}{\alpha}m_{\delta}''(x)+\Bigg[\frac{(\lambda+\delta - r)}{\alpha r}-(x-x^{*})\Bigg]m_{\delta}'(x)+\frac{\delta}{r}m_{\delta}(x)=0, \qquad x\geq x^{*}. \label{Appendix A: Mathematical Proofs-Equation4}
\end{align}

\vspace{0.3cm}

Letting $f(y):=m_{\delta}(x)$, such that $y$ is associated with the change of variable $y:=y(x)=-\alpha (x-x^{*})$, \eqref{Appendix A: Mathematical Proofs-Equation4} reduces to Kummer\rq s Confluent Hypergeometric Equation \citep{Book:Slater1960}

\vspace{0.3cm}

\begin{align}
    y\cdot f''(y) + (c-y) f'(y) - a f(y) =0, \qquad y<0,
    \label{Appendix A: Mathematical Proofs-Equation5}
\end{align}

\vspace{0.3cm}

for $a=-\frac{\delta}{r}$ and $c=1-\frac{\lambda + \delta}{r}$, with regular singular point at $y=0$ and irregular singular point at $y=-\infty$ (corresponding to $x=x^{*}$ and $x=\infty$, respectively). A general solution of \eqref{Appendix A: Mathematical Proofs-Equation5} is given by

\vspace{0.3cm}

\begin{align}
    m_{\delta}(x)=f(y)= \begin{cases} 1 \hspace{9.5cm} x < x^{*},\\
    A_{1}M\left(-\frac{\delta}{r},1-\frac{\lambda+\delta}{r};y(x)\right)+A_{2}e^{y(x)}U\left(1-\frac{\lambda}{r}, 1-\frac{\lambda+\delta}{r};-y(x)\right) \hspace{0.2cm} x \ge x^{*},
    \end{cases}\\
    \label{Appendix A: Mathematical Proofs-Equation6}
\end{align}
\normalsize

\vspace{0.3cm}

for arbitrary constants $A_{1},A_{2} \in \mathbb{R}$. Here, 

\vspace{0.3cm}

\begin{align}
    M(a, c; z)={ }_{1} F_{1}(a, c; z)=\sum_{n=0}^{\infty} \frac{(a)_{n}}{(c)_{n}} \frac{z^{n}}{n !}
    \label{Appendix A: Mathematical Proofs-Equation7}
\end{align}

\vspace{0.3cm}

is Kummer\rq s Confluent Hypergeometric Function \citep{Article:Kummer1837} and $(a)_{n}=\frac{\Gamma(a+n)}{\Gamma(n)}$ denotes the Pochhammer symbol \citep{Book:Seaborn1991}. In a similar manner,

\vspace{0.3cm}

\begin{align}
    U(a, c; z)=\left\{\begin{array}{ll}\frac{\Gamma(1-c)}{\Gamma(1+a-c)} M(a, c; z)+\frac{\Gamma(c-1)}{\Gamma(a)} z^{1-c} M(1+a-c, 2-c; z) & c \notin \mathbb{Z}, \\ \lim\limits_{\theta \rightarrow c} U(a, \theta; z) & c \in \mathbb{Z},\end{array}\right.
    \label{Appendix A: Mathematical Proofs-Equation8}
\end{align}

\vspace{0.3cm}

is Tricomi\rq s Confluent Hypergeometric Function \citep{Article:Tricomi1947}. This function is generally complex-valued when its argument $z$ is negative, i.e. when $x \ge x^{*}$ in the case of interest. We seek a real-valued solution of $m_{\delta}(x)$ over the entire domain, therefore an alternative independent pair of solutions, here, $M(a,c;z)$ and $e^{z}U(c-a,c;-z)$, to \eqref{Appendix A: Mathematical Proofs-Equation5} are chosen for $x \ge x^{*}$.

To determine the constants $A_1$ and $A_2$ we consider the boundary conditions for $m_{\delta}(x)$ at $x^*$ and at infinity. 
Applying equation (13.1.27) of \cite{Book:Abramowitz1964}, also known as Kummer\rq s Transformation $M(a, c; z) = e^{z} M(c-a, c;-z)$, we write \eqref{Appendix A: Mathematical Proofs-Equation6} such that

\vspace{0.3cm}

\begin{align}
    m_{\delta}(x)=
    e^{y(x)}\left[A_{1}M\left(1-\frac{\lambda}{r},1-\frac{\lambda+\delta}{r};-y(x)\right)+A_{2}U\left(1-\frac{\lambda}{r}, 1-\frac{\lambda+\delta}{r};-y(x)\right)\right],\\
   \label{Appendix A: Mathematical Proofs-Equation9}
\end{align}

\vspace{0.3cm}

for $x \ge x^{*}$. For $z \rightarrow \infty$, it is well-known that

\vspace{0.3cm}

\begin{align}
    M(a, c; z)=\frac{\Gamma(c)}{\Gamma(a)} e^{z} z^{a-c}\left[1+O\left(|z|^{-1}\right)\right]
    \label{Appendix A: Mathematical Proofs-Equation10}
\end{align}

\vspace{0.3cm}

and

\vspace{0.3cm}

\begin{align}
    U(a, c; z)= z^{-a}\left[1+O\left(|z|^{-1}\right)\right]
    \label{Appendix A: Mathematical Proofs-Equation11}
\end{align}

\vspace{0.3cm}

(see for example, equations (13.1.4) and (13.1.8) of \cite{Book:Abramowitz1964}). Asymptotic behaviours of the first and second terms of \eqref{Appendix A: Mathematical Proofs-Equation9} as $y(x) \rightarrow -\infty$ are therefore given by

\vspace{0.3cm}

\begin{align}
    \frac{\Gamma\left(1-\frac{\lambda+\delta}{r}\right)}{\Gamma\left(1-\frac{\lambda}{r}\right)}\left(-y(x)\right)^{\frac{\delta}{r}}\left(1+O\left(|-y(x)|^{-1}\right)\right)
    \label{Appendix A: Mathematical Proofs-Equation12}
\end{align}

\vspace{0.3cm}

and

\vspace{0.3cm}

\begin{align}
    e^{y(x)}\left(-y(x)\right)^{\frac{\lambda}{r}-1}\left(1+O\left(|-y(x)|^{-1}\right)\right),
    \label{Appendix A: Mathematical Proofs-Equation13}
\end{align}

\vspace{0.3cm}

respectively. For $x \rightarrow \infty$, \eqref{Appendix A: Mathematical Proofs-Equation12} is unbounded, while \eqref{Appendix A: Mathematical Proofs-Equation13} tends to zero. The boundary condition $\lim\limits_{x\to\infty} m_{\delta}(x) = 0$, by definition of $m_{\delta}(x)$ in \eqref{TheTrappingTime-Section3-Equation2}, thus implies
that $A_{1}=0$. Letting $x=x^{*}$ in \eqref{Appendix A: Mathematical Proofs-Equation3} and \eqref{Appendix A: Mathematical Proofs-Equation6} yields

\vspace{0.3cm}

\begin{align}
    \frac{\lambda}{(\lambda + \delta)}=A_{2}U\left(1-\frac{\lambda}{r}, 1-\frac{\lambda+\delta}{r};0\right).
    \label{Appendix A: Mathematical Proofs-Equation14}
\end{align}

\vspace{0.3cm}

Hence, $A_{2}=\frac{\lambda}{(\lambda + \delta)U\left(1-\frac{\lambda}{r}, 1-\frac{\lambda+\delta}{r};0\right)}$ and the Laplace transform of the trapping time for $x\geq x^*$ is given by \eqref{TheTrappingTime-Section3-Equation3}.

 \vspace{0.3cm}
 
\subsubsection{Proof of Corollary \ref{TheTrappingTime-Section3-Corollary1}} \label{ProofofCorollary3.1}

We differentiate Tricomi\rq s Confluent Hypergeometric Function \eqref{Appendix A: Mathematical Proofs-Equation8} with respect to its second parameter. Denote

    \vspace{0.3cm}

    \begin{align}
        U^{(c)}(a, c; z)\equiv \frac{d}{d c} U(a, c; z).
        \label{Appendix A: Mathematical Proofs-Equation15}
    \end{align}
    
    \vspace{0.3cm}

    A closed form expression of the aforementioned derivative can be given in terms of series expansions, such that

    \vspace{0.3cm}

    \begin{align}
        \begin{split}
            U^{(c)}(a, c; z)&=(\eta(a-c+1)-\pi \cot (c\pi)) U(a, c; z)\\
            &-\frac{\Gamma(c-1) z^{1-c} \log (z)}{\Gamma(a)}{ }M(a-c+1 , 2-c ; z)\\&- 
            \frac{\Gamma(c-1) z^{1-c}}{\Gamma(a)} \sum_{k=0}^{\infty} \frac{(a-c+1)_{k}(\eta(a-c+k+1)-\eta(2-c+k)) z^{k}}{(2-c)_{k} k !}\\
            &-\frac{\Gamma(1-c)}{\Gamma(a-c+1)} \sum_{k=0}^{\infty} \frac{\eta(c+k)(a)_{k} z^{k}}{(c)_{k} k !}, \qquad c \notin \mathbb{Z},
        \end{split}
        \label{Appendix A: Mathematical Proofs-Equation16}
    \end{align}
    
    \vspace{0.3cm}

    where $\eta(z)=\frac{d \ln\left[\Gamma(z)\right]}{dz}=\frac{\Gamma '(z)}{\Gamma(z)}$ corresponds to equation (6.3.1) of \cite{Book:Abramowitz1964}, also known as the digamma function. Calculating \eqref{TheTrappingTime-Section3-Equation8} and using \eqref{Appendix A: Mathematical Proofs-Equation16}, one can derive the expected trapping time \eqref{TheTrappingTime-Section3-Equation9}.
    
\vspace{0.3cm}

\subsubsection{Proof of Proposition \ref{CostofSocialProtection-Subsection52-Proposition1}} \label{ProofofProposition5.1}

Since 

    \begin{align}
S=\frac{\beta}{\delta} \left[1-e^{-\delta \tau^{\scaleto{\pi^{*}(\kappa,\theta)}{5pt}}_{x}}\right],
        \label{Appendix A: Mathematical Proofs-Equation17}
    \end{align}
    
then we consider $m^{\scaleto{\pi^{*}(\kappa,\theta)}{5pt}}_\delta(x)$, the Laplace transform for the insured process obtained in \eqref{IntroducingMicroinsurance-Section4-Equation2} with capital growth $r^{\scaleto{\pi^{*}(\kappa,\theta)}{5pt}}$ to compute $V^{\scaleto{\pi^{*}(\kappa,\theta)}{5pt}}(x)$ when capital losses are exponentially distributed with parameter $\alpha^{\scaleto{(\kappa)}{5pt}}>0$. This yields \eqref{CostofSocialProtection-Subsection52-Equation3}.

\vspace{0.3cm}

\subsubsection{Proof of Proposition \ref{CostofSocialProtection-Subsection52-Proposition2}} \label{ProofofProposition5.2}

Following a similar procedure to that in Proposition \ref{TheTrappingTime-Section3-Proposition1}, consider the integral

\vspace{0.3cm}

\begin{align}
A(x)&:=\int_{x-x^{\scaleto{\pi^{*}(\kappa,\theta)*}{5pt}}}^{\infty} w(x-x^{\scaleto{\pi^{*}(\kappa,\theta)*}{5pt}}, z-(x-x^{\scaleto{\pi^{*}(\kappa,\theta)*}{5pt}}))dG_{Z}(z) \\ \\ 
&= \int_{x-x^{\scaleto{\pi^{*}(\kappa,\theta)*}{5pt}}}^{\infty} \left[z - (x - x^{\scaleto{\pi^{*}(\kappa,\theta)*}{5pt}}) + M^{\scaleto{(\kappa)}{5pt}}-x^{\scaleto{\pi^{*}(\kappa,\theta)*}{5pt}}\right] \alpha^{\scaleto{(\kappa)}{5pt}} e^{-\alpha^{\scaleto{(\kappa)}{5pt}} z} dz \\ \\ 
&= \left(\frac{1}{\alpha^{\scaleto{(\kappa)}{5pt}}} + M^{\scaleto{(\kappa)}{5pt}}-x^{\scaleto{\pi^{*}(\kappa,\theta)*}{5pt}}\right)e^{-\alpha^{\scaleto{(\kappa)}{5pt}} (x-x^{\scaleto{\pi^{*}(\kappa,\theta)*}{5pt}})},\label{Appendix A: Mathematical Proofs-Equation18}
\end{align}

\vspace{0.3cm}

which under the assumption $w(x_{1},x_{2})= x_{2} + M^{\scaleto{(\kappa)}{5pt}}-x^{\scaleto{\pi^{*}(\kappa,\theta)*}{5pt}}$ yields a modified version of the IDE \eqref{Appendix A: Mathematical Proofs-Equation1} given by

\vspace{0.3cm}

\begin{equation}
\begin{aligned}
    r^{\scaleto{\pi^{*}(\kappa,\theta)}{5pt}}(x-x^{\scaleto{\pi^{*}(\kappa,\theta)*}{5pt}})m^{\scaleto{\pi^{*}(\kappa,\theta)\prime}{5pt}}_{\delta,w}(x)-(\lambda + \delta) m^{\scaleto{\pi^{*}(\kappa,\theta)}{5pt}}_{\delta,w}(x)+\lambda \int_{0}^{x-x^{\scaleto{\pi^{*}(\kappa,\theta)*}{5pt}}}m^{\scaleto{\pi^{*}(\kappa,\theta)}{5pt}}_{\delta,w}(x-z)\alpha^{\scaleto{(\kappa)}{5pt}} e^{-\alpha^{\scaleto{(\kappa)}{5pt}} z} dz \\ \\ 
    =-\lambda \left(\frac{1}{\alpha^{\scaleto{(\kappa)}{5pt}}} + M^{\scaleto{(\kappa)}{5pt}}-x^{\scaleto{\pi^{*}(\kappa,\theta)*}{5pt}}\right)e^{-\alpha^{\scaleto{(\kappa)}{5pt}} (x-x^{\scaleto{\pi^{*}(\kappa,\theta)*}{5pt}})} , \hspace{0.1cm} x \ge x^{\scaleto{\pi^{*}(\kappa,\theta)*}{5pt}}.
    \label{Appendix A: Mathematical Proofs-Equation19}
\end{aligned}
\end{equation}
\normalsize

\vspace{0.3cm}

Solving \eqref{Appendix A: Mathematical Proofs-Equation19} in the same manner as \eqref{Appendix A: Mathematical Proofs-Equation3} gives \eqref{CostofSocialProtection-Subsection52-Equation5}.

\vspace{0.3cm}

\subsubsection{Proof of Proposition \ref{GeneralSetting-Subsection61-Proposition1}} \label{ProofofProposition6.1}

Under the alternative microinsurance subsidy scheme, the Laplace transform of the trapping time satisfies the following differential equations:

\vspace{0.3cm}

\small
\begin{align}
0 =
\begin{cases}                 -\frac{\left(x-x^{\scaleto{(\mathcal{A})*}{5pt}}\right)}{\alpha^{\scaleto{(\kappa)}{5pt}}} m_{\delta}^{\scaleto{(\mathcal{A})}{5pt}\prime \prime}(x)+\left[\frac{(\lambda+\delta-r)}{\alpha^{\scaleto{(\kappa)}{5pt}} r}-\left(x-x^{\scaleto{(\mathcal{A})*}{5pt}}\right)\right] m_{\delta}^{\scaleto{(\mathcal{A})}{5pt} \prime}(x)+\frac{\delta}{r} m_{\delta}^{\scaleto{(\mathcal{A})}{5pt}}(x) \hspace{1cm} \textit{for $ x^{\scaleto{(\mathcal{A})*}{5pt}} \leq x \leq B$}, \\                -\frac{\left(x-x^{\scaleto{(\mathcal{A})*}{5pt}}\right)}{\alpha^{\scaleto{(\kappa)}{5pt}}} m_{\delta}^{\scaleto{(\mathcal{A})}{5pt}\prime \prime}(x)+\left[\frac{(\lambda+\delta-r^{\scaleto{(\kappa)}{5pt}})}{\alpha^{\scaleto{(\kappa)}{5pt}} r^{\scaleto{(\kappa)}{5pt}}}-\left(x-x^{\scaleto{(\mathcal{A})*}{5pt}}\right)\right] m_{\delta}^{\scaleto{(\mathcal{A})}{5pt}\prime}(x)+\frac{\delta}{r^{\scaleto{(\kappa)}{5pt}}} m_{\delta}^{\scaleto{(\mathcal{A})}{5pt}}(x) \hspace{0.3cm} \textit{for  $x \geq B$}.
\end{cases}
\label{Appendix A: Mathematical Proofs-Equation20}
\end{align}
\normalsize

\vspace{0.3cm}

As in Proposition \ref{TheTrappingTime-Section3-Proposition1}, use of the change of variable  $y^{\scaleto{(\mathcal{A})}{5pt}}:=y^{\scaleto{(\mathcal{A})}{5pt}}(x)=-\alpha^{\scaleto{(\kappa)}{5pt}} (x-x^{\scaleto{(\mathcal{A})*}{5pt}})$ leads to Kummer\rq s Confluent Hypergeometric Equation, thus equation \eqref{GeneralSetting-Subsection61-Equation1} is obtained for arbitrary constants $C_{1},C_{2},C_{3},C_{4} \in \mathbb{R}$. Under the boundary condition $\lim\limits_{x\to\infty} m_{\delta}^{\scaleto{(\mathcal{A})}{5pt}}(x) = 0$ with asymptotic behaviour of the Kummer function $M(a,c;z)$ as presented in Proposition \ref{TheTrappingTime-Section3-Proposition1}, we deduce that 

\vspace{0.3cm}

\begin{align}
C_{3}=0.\label{Appendix A: Mathematical Proofs-Equation21}
\end{align}

\vspace{0.3cm}

Then, since $m_{\delta}^{\scaleto{(\mathcal{A})}{5pt}}(x^{\scaleto{(\mathcal{A})*}{5pt}})=\frac{\lambda}{\lambda + \delta}$, we obtain

\vspace{0.3cm}

\begin{align}
C_{1}=\frac{\lambda}{\lambda + \delta} - C_{2}U\left(1-\frac{\lambda}{r}, 1-\frac{\lambda+\delta}{r};0\right).\label{Appendix A: Mathematical Proofs-Equation22}
\end{align}

\vspace{0.3cm}

Due to the continuity of the functions $ m_{\delta}^{\scaleto{(\mathcal{A})}{5pt}}(x)$ and  $m_{\delta}^{\scaleto{(\mathcal{A})}{5pt}\prime}(x)$ at $x=B$ and the differential properties of the Confluent Hypergeometric Functions:

\vspace{0.3cm}

\begin{align}
    \frac{d}{dz}M(a,c;z)=\frac{a}{c}M(a+1,c+1;z),
    \label{Appendix A: Mathematical Proofs-Equation23}
\end{align}

\begin{align}
    \frac{d}{dz}U(a,c;z)=-aU(a+1,c+1;z),
    \label{Appendix A: Mathematical Proofs-Equation24}
\end{align}

\vspace{0.3cm}

upon simplification,

\vspace{0.3cm}

\scriptsize
\begin{align}
    \begin{split}
        C_{4}=\frac{\left[\frac{\lambda}{\lambda + \delta} - C_{2}U\left(1-\frac{\lambda}{r}, 1-\frac{\lambda+\delta}{r};0\right)\right]M\left(-\frac{\delta}{r},1-\frac{\lambda+\delta}{r};y^{\scaleto{(\mathcal{A})}{5pt}}(B)\right)
        +C_{2}e^{y^{\scaleto{(\mathcal{A})}{5pt}}(B)}U\left(1-\frac{\lambda}{r}, 1-\frac{\lambda+\delta}{r};-y^{\scaleto{(\mathcal{A})}{5pt}}(B)\right)}{e^{y^{\scaleto{(\mathcal{A})}{5pt}}(B)}U\left(1-\frac{\lambda}{r^{\scaleto{(\kappa)}{5pt}}}, 1-\frac{\lambda+\delta}{r^{\scaleto{(\kappa)}{5pt}}};-y^{\scaleto{(\mathcal{A})}{5pt}}(B)\right)}
    \end{split}\\
    \label{Appendix A: Mathematical Proofs-Equation25}
\end{align}
\normalsize 

\vspace{0.3cm}

and 

\vspace{0.3cm}

\begin{align}
    \begin{split}
        C_{2}=\frac{\frac{\lambda}{\lambda+\delta}\left[\frac{\delta \alpha^{\scaleto{(\kappa)}{5pt}}}{(r-\lambda-\delta)}M\left(1-\frac{\delta}{r}, 2-\frac{\lambda+\delta}{r};y^{\scaleto{(\mathcal{A})}{5pt}}(B)\right)+M\left(-\frac{\delta}{r}, 1-\frac{\lambda+\delta}{r};y^{\scaleto{(\mathcal{A})}{5pt}}(B)\right)\left(\alpha^{\scaleto{(\kappa)}{5pt}} - D\right)\right]}{K},
    \end{split}\\
    \label{Appendix A: Mathematical Proofs-Equation26}
\end{align}

\vspace{0.3cm}

where 

\vspace{0.3cm}

\begin{align}
    D:=\frac{\alpha^{\scaleto{(\kappa)}{5pt}}\left(\frac{\lambda}{r^{\scaleto{(\kappa)}{5pt}}}-1\right)U\left(2-\frac{\lambda}{r^{\scaleto{(\kappa)}{5pt}}},2-\frac{\lambda+\delta}{r^{\scaleto{(\kappa)}{5pt}}};-y^{\scaleto{(\mathcal{A})}{5pt}}(B)\right)}{U\left(1-\frac{\lambda}{r^{\scaleto{(\kappa)}{5pt}}},1-\frac{\lambda+\delta}{r^{\scaleto{(\kappa)}{5pt}}};-y^{\scaleto{(\mathcal{A})}{5pt}}(B)\right)}
    \label{Appendix A: Mathematical Proofs-Equation27}
\end{align}

\vspace{0.3cm}

and 

\vspace{0.3cm}

\small
\begin{align}
\begin{split}
K: &= M\left(-\frac{\delta}{r}, 1-\frac{\lambda+\delta}{r};y^{\scaleto{(\mathcal{A})}{5pt}}(B)\right)U\left(1-\frac{\lambda}{r}, 1-\frac{\lambda+\delta}{r};0\right)\left(\alpha^{\scaleto{(\kappa)}{5pt}}-D\right)
\\
&+D e^{y^{\scaleto{(\mathcal{A})}{5pt}}(B)}U\left(1-\frac{\lambda}{r}, 1-\frac{\lambda+\delta}{r};-y^{\scaleto{(\mathcal{A})}{5pt}}(B)\right) \\ 
&+\frac{\delta \alpha^{\scaleto{(\kappa)}{5pt}}}{(r-\lambda -\delta)}M\left(1-\frac{\delta}{r}, 2-\frac{\lambda+\delta}{r};y^{\scaleto{(\mathcal{A})}{5pt}}(B)\right)U\left(1-\frac{\lambda}{r}, 1-\frac{\lambda+\delta}{r};0\right)
\\
&-\alpha^{\scaleto{(\kappa)}{5pt}}e^{y^{\scaleto{(\mathcal{A})}{5pt}}(B)}\left(\frac{\lambda}{r}-1\right)U\left(2-\frac{\lambda}{r}, 2-\frac{\lambda+\delta}{r};-y^{\scaleto{(\mathcal{A})}{5pt}}(B)\right).
\label{Appendix A: Mathematical Proofs-Equation28}
\end{split}
\end{align}
\normalsize

\subsection{Effects of Selected Underlying Factors on the Trapping Probability and the Cost of Social Protection}\label{Appendix B: Effects of Underlying Factors on the Trapping Probability and the Cost of Social Protection}

For selected parameters, we consider their influence on the trapping probability and the cost of social protection by varying them in a reasonable range, keeping all other parameters constant. The reference setup is given below.

\textbf{Reference setup:} $Z_{i}\sim Exp(1)$, $a = 0.1$, $b = 1.4$, $c = 0.4$, $\lambda=1$, $x^{*} = x^{\scaleto{(\kappa)*}{5pt}} = x^{\scaleto{\pi(\kappa, \theta)*}{5pt}} = x^{\scaleto{(\mathcal{A})*}{5pt}} = 1$, $\kappa = 0.5$, $\theta = 0.5$, $\delta=0.1$ and $\epsilon=0.01$.

\begin{figure}[p]
	\begin{center}
	\hspace{-5.3cm}
	\begin{subfigure}[b]{0.5\linewidth}
  	\hspace{0.95cm} \includegraphics[width=4.1cm, height=1.5cm]{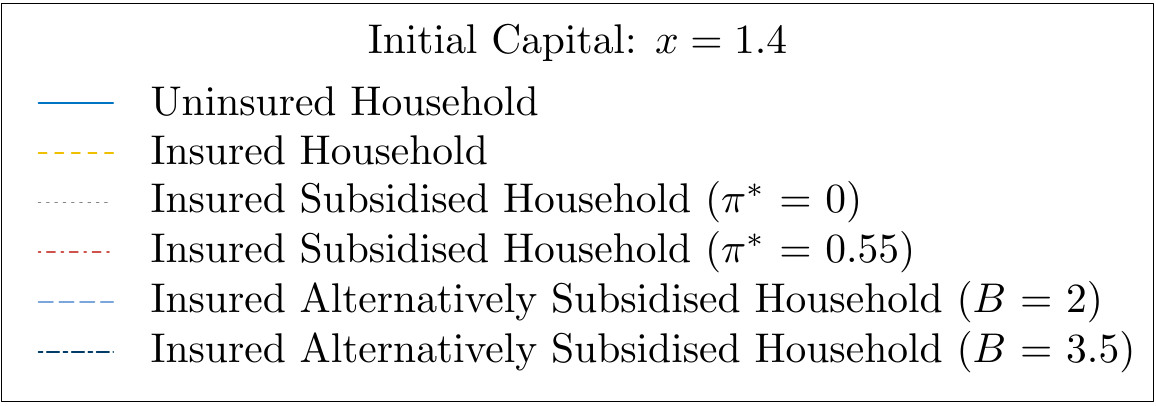}
	\end{subfigure}
	\end{center}
  	\centering
  	\includegraphics[width=5.5cm, height=5.5cm]{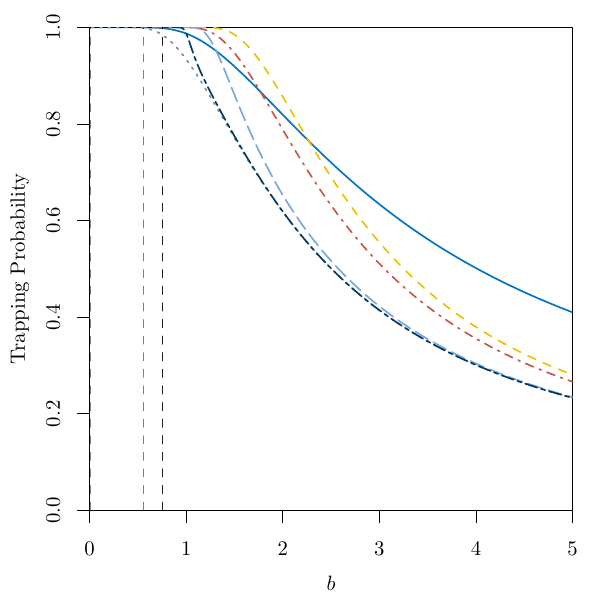}
 	 \hspace{1cm}
 	 \includegraphics[width=5.5cm, height=5.5cm]{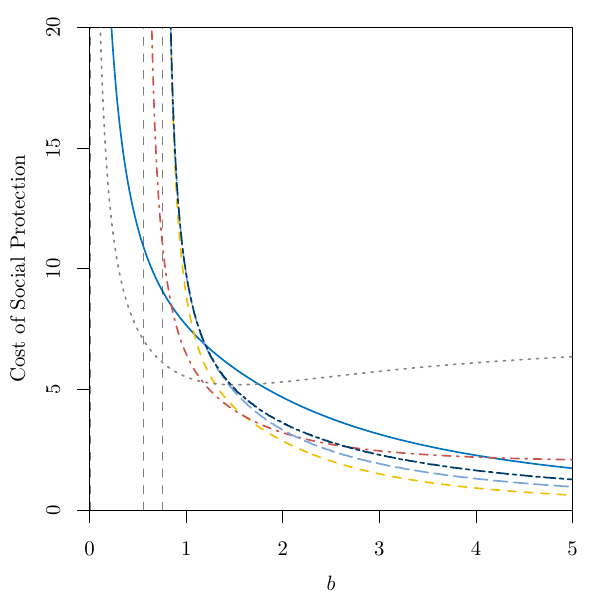}
  	\vspace{0.1cm}
 	 \includegraphics[width=5.5cm, height=5.5cm]{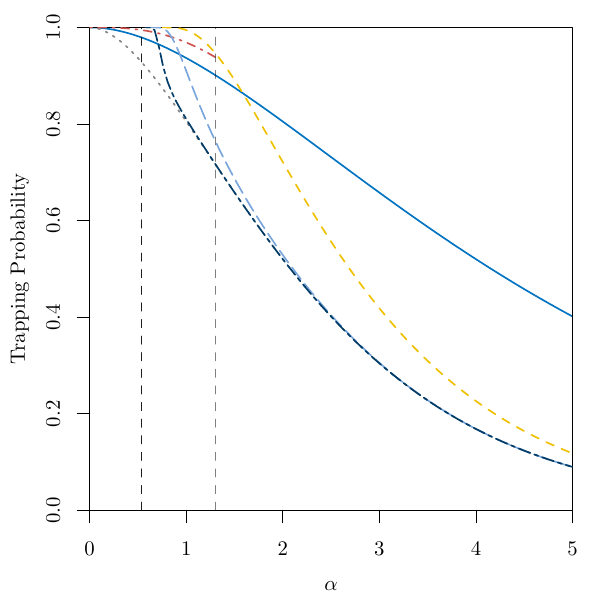}
 	 \hspace{1cm}
 	 \includegraphics[width=5.5cm, height=5.5cm]{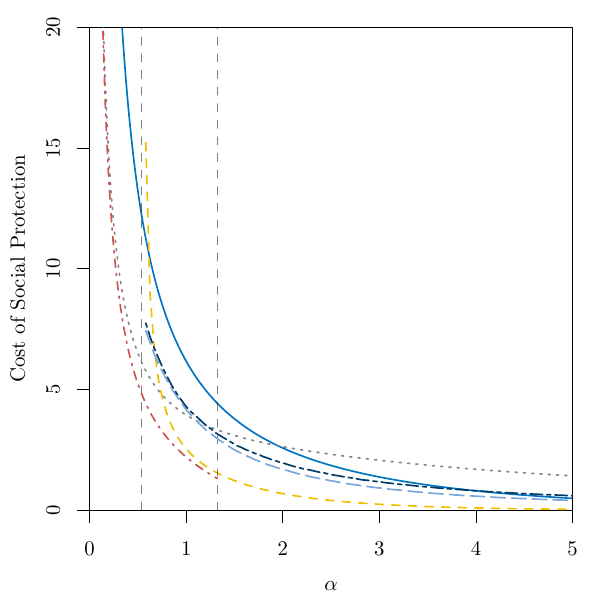}
  	\vspace{0.1cm}
  	\includegraphics[width=5.5cm, height=5.5cm]{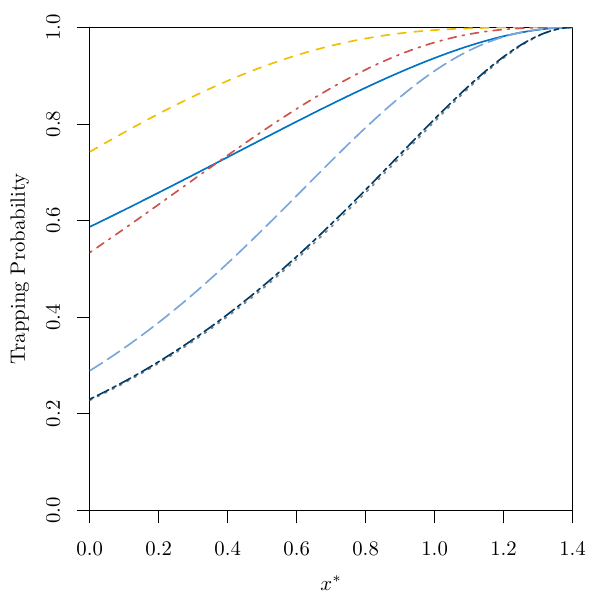}
  	\hspace{1cm}
  	\includegraphics[width=5.5cm, height=5.5cm]{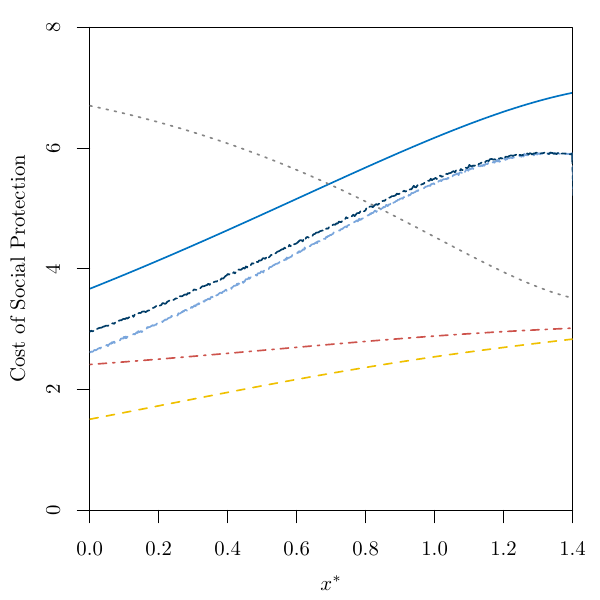}
  \caption{Effects of the rate of income generation $(0<b)$, the parameter of the exponential distribution $(\alpha > 0)$ (i.e., expected capital loss size) and the critical capital $(x \geq x^{*})$ for initial capital $x=1.4$.}
  \label{Appendix B: Effects of Underlying Factors on the Trapping Probability and the Cost of Social Protection-Figure1}
\end{figure}

\begin{figure}[p]
	\begin{center}
	\hspace{-5.3cm}
	\begin{subfigure}[b]{0.5\linewidth}
  	\hspace{0.95cm} \includegraphics[width=4.1cm, height=1.5cm]{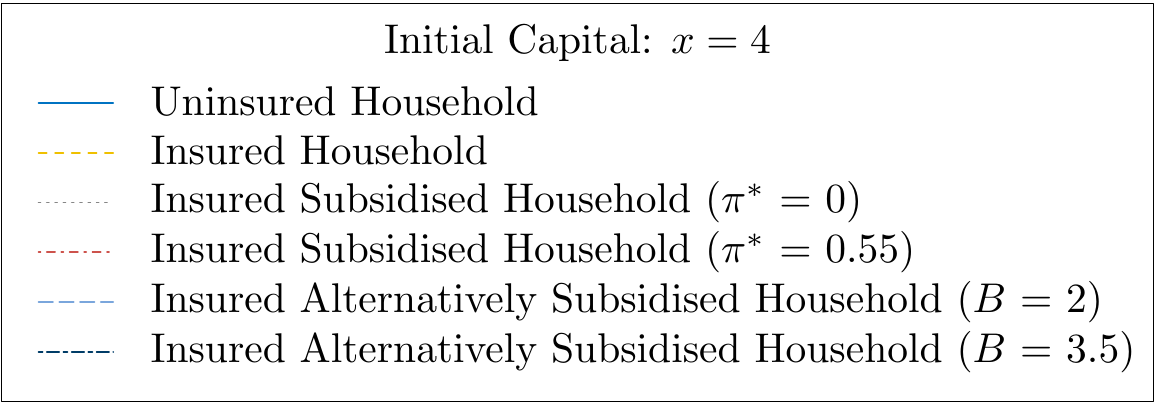}
	\end{subfigure}
	\end{center}
  	\centering
  	\includegraphics[width=5.5cm, height=5.5cm]{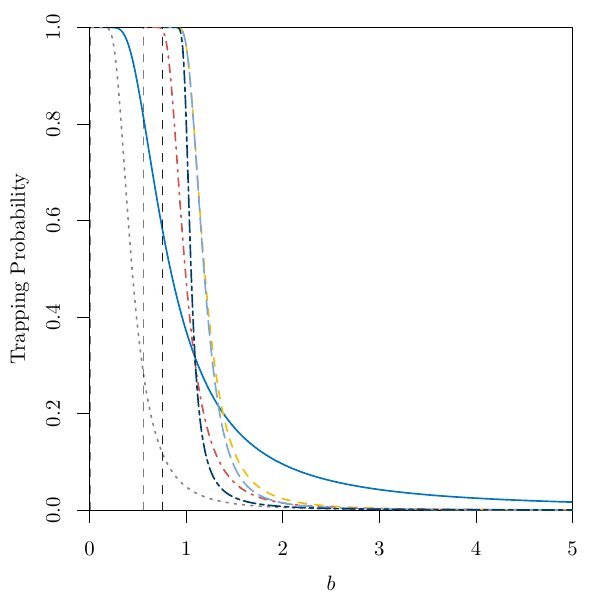}
 	 \hspace{1cm}
 	 \includegraphics[width=5.5cm, height=5.5cm]{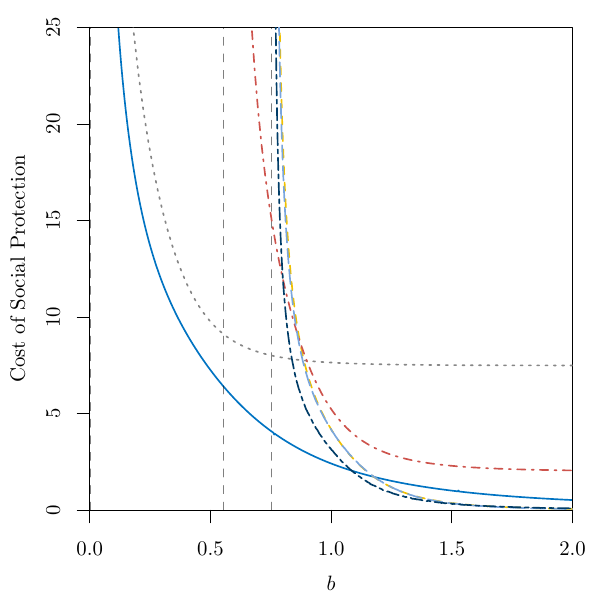}
  	\vspace{0.1cm}
 	 \includegraphics[width=5.5cm, height=5.5cm]{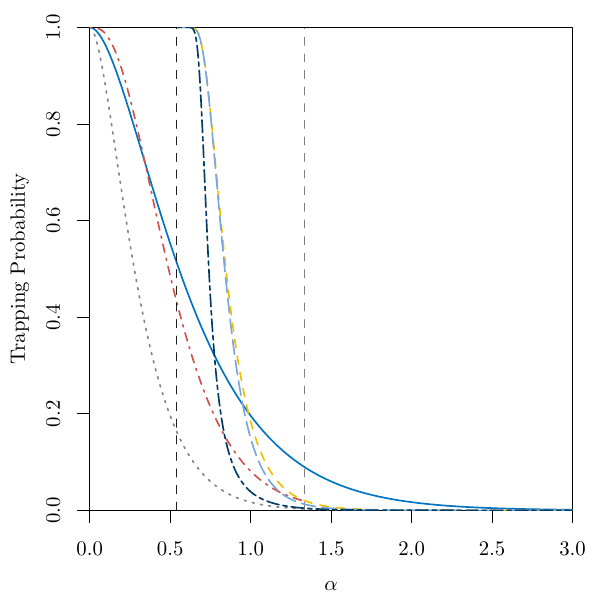}
 	 \hspace{1cm}
 	 \includegraphics[width=5.5cm, height=5.5cm]{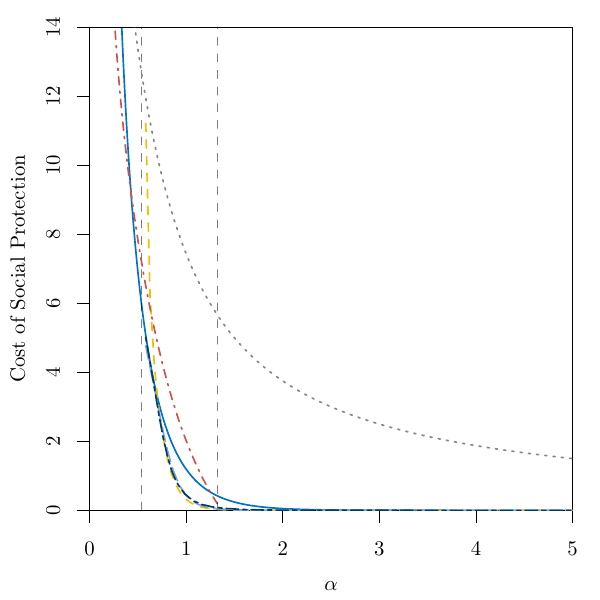}
  	\vspace{0.1cm}
  	\includegraphics[width=5.5cm, height=5.5cm]{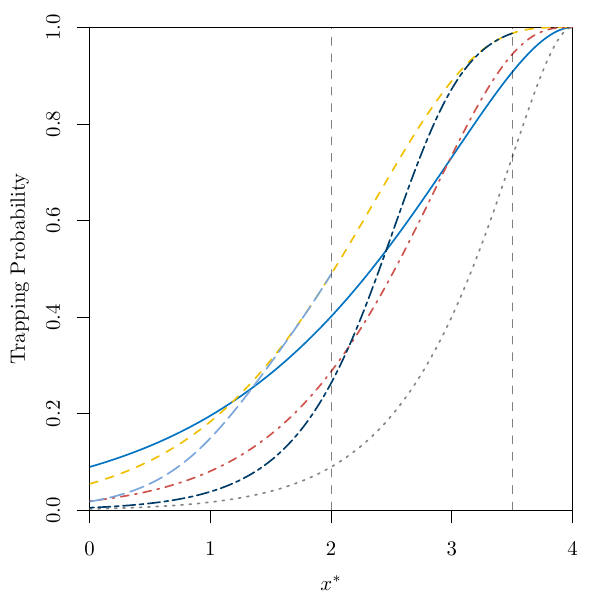}
  	\hspace{1cm}
  	\includegraphics[width=5.5cm, height=5.5cm]{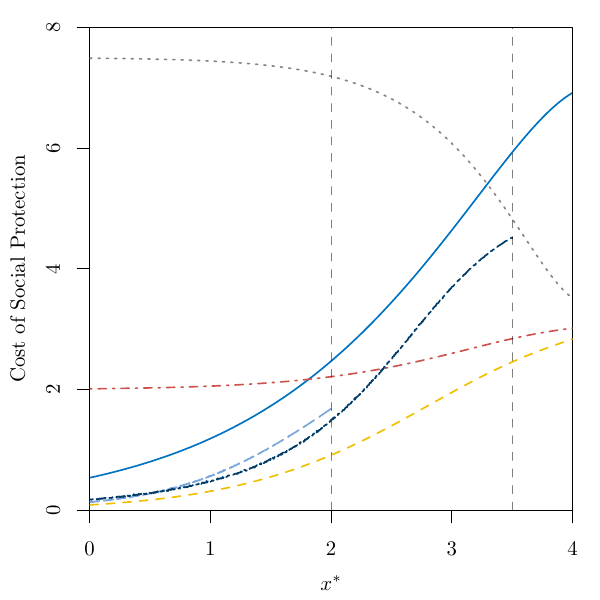}
  \caption{Effects of the rate of income generation $(0<b)$, the parameter of the exponential distribution $(\alpha > 0)$ (i.e., expected capital loss size) and the critical capital $(x \geq x^{*})$ for initial capital $x=4$.}
  \label{Appendix B: Effects of Underlying Factors on the Trapping Probability and the Cost of Social Protection-Figure2}
\end{figure}

\begin{figure}[p]
	\begin{center}
	\hspace{-5.3cm}
	\begin{subfigure}[b]{0.5\linewidth}
  	\hspace{0.95cm} \includegraphics[width=4.1cm, height=1.5cm]{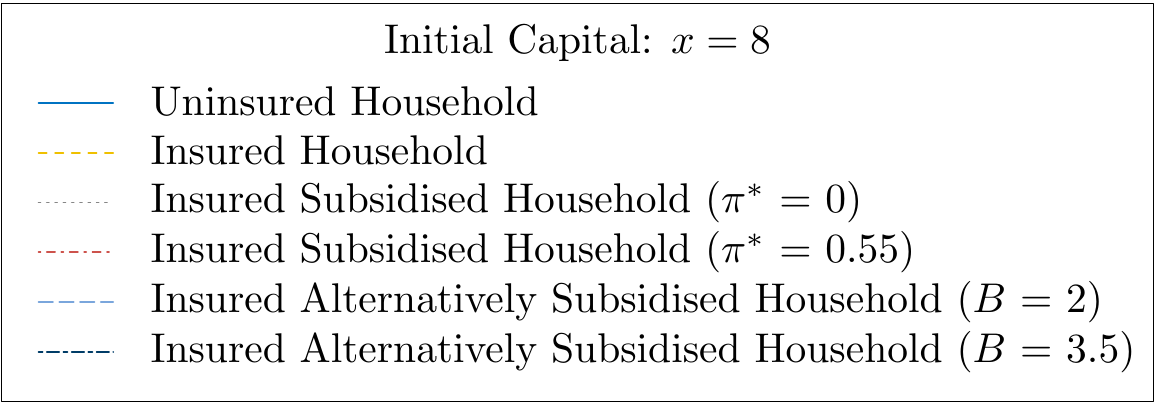}
	\end{subfigure}
	\end{center}
  	\centering
  	\includegraphics[width=5.5cm, height=5.5cm]{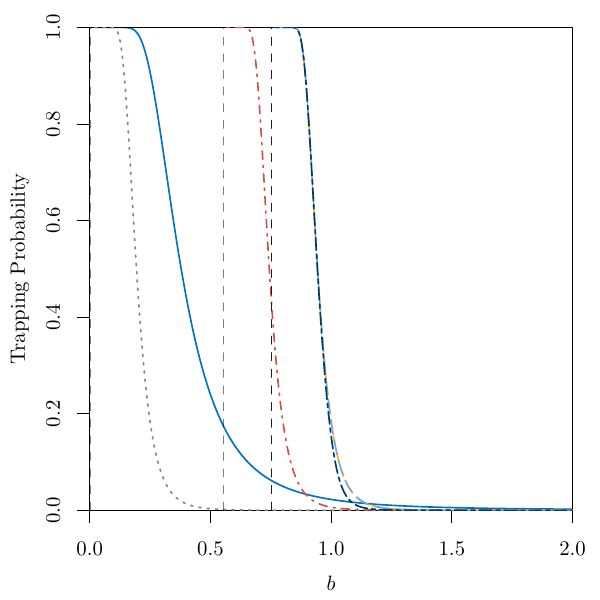}
 	 \hspace{1cm}
 	 \includegraphics[width=5.5cm, height=5.5cm]{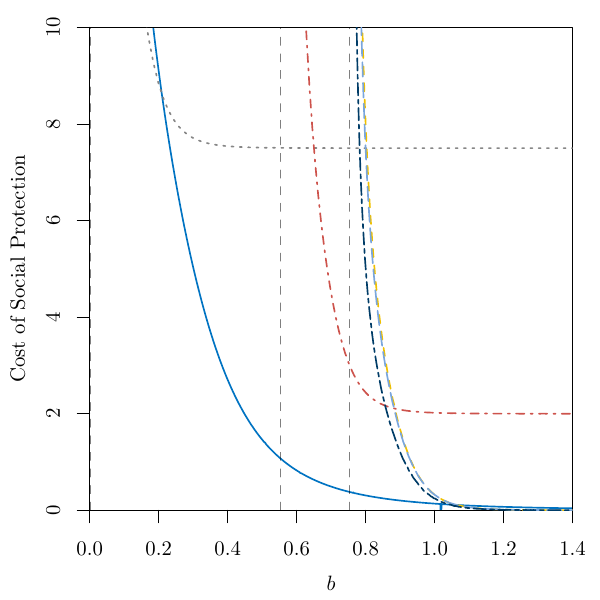}
  	\vspace{0.1cm}
 	 \includegraphics[width=5.5cm, height=5.5cm]{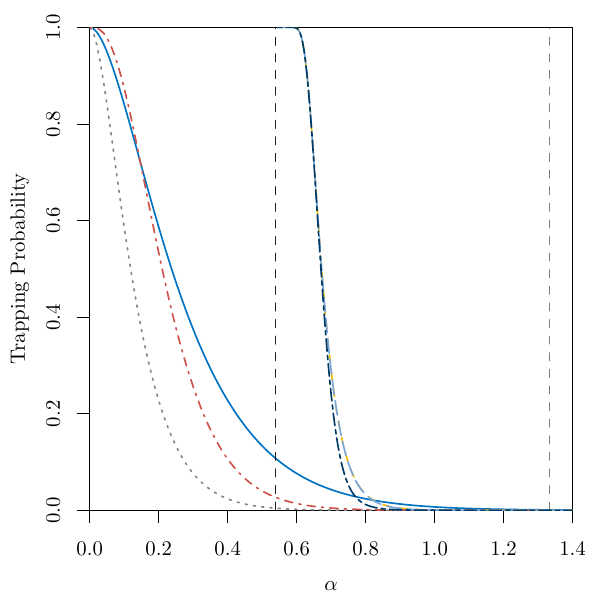}
 	 \hspace{1cm}
 	 \includegraphics[width=5.5cm, height=5.5cm]{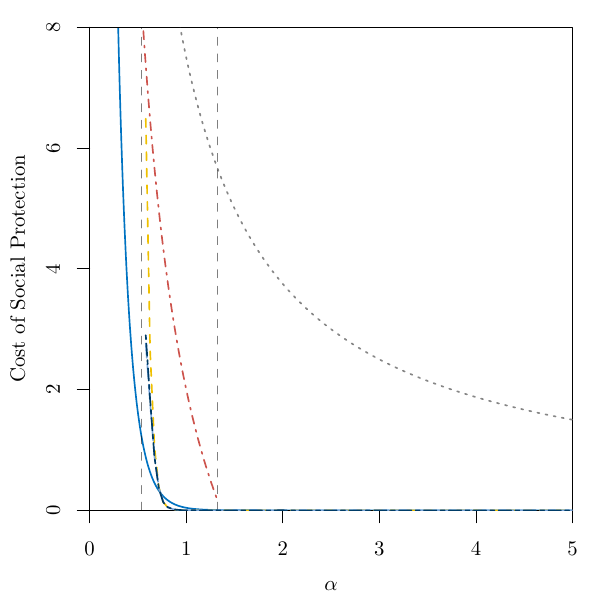}
  	\vspace{0.1cm}
  	\includegraphics[width=5.5cm, height=5.5cm]{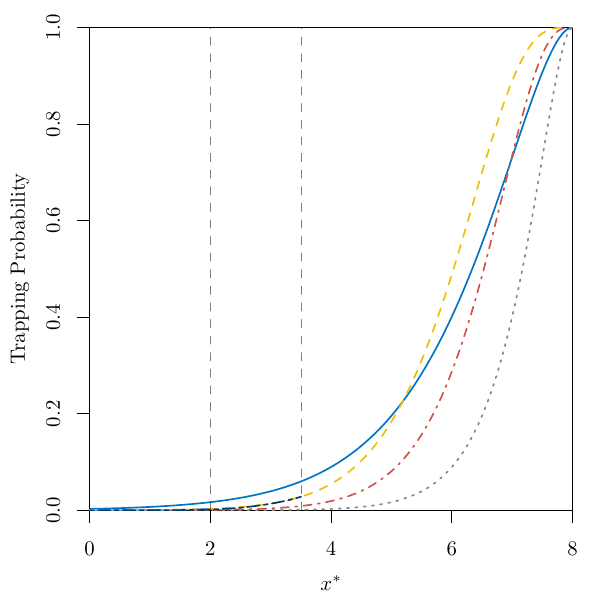}
  	\hspace{1cm}
  	\includegraphics[width=5.5cm, height=5.5cm]{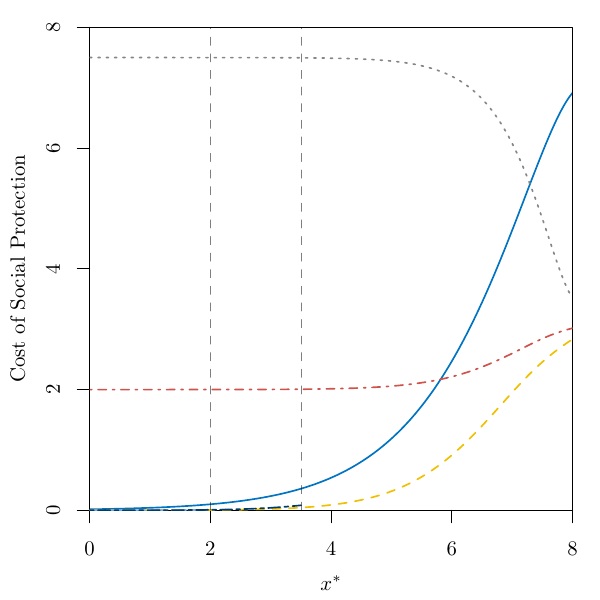}
  \caption{Effects of the rate of income generation $(0<b)$, the parameter of the exponential distribution $(\alpha > 0)$ (i.e., expected capital loss size) and the critical capital $(x \geq x^{*})$ for initial capital $x=8$.}
  \label{Appendix B: Effects of Underlying Factors on the Trapping Probability and the Cost of Social Protection-Figure3}
\end{figure}

Figures \ref{Appendix B: Effects of Underlying Factors on the Trapping Probability and the Cost of Social Protection-Figure1}, \ref{Appendix B: Effects of Underlying Factors on the Trapping Probability and the Cost of Social Protection-Figure2} and \ref{Appendix B: Effects of Underlying Factors on the Trapping Probability and the Cost of Social Protection-Figure3} show that, when the capital growth rate $r$ is close to zero (this could occur in two situations, when $b$ is close to zero, for an uninsured household, or, when the premiums $\pi < b$ or $\pi^{*} < b$, depending the case, have values close to $b$, for an insured household), both the trapping probability and the cost of social protection increase for the three groups of households with different levels of initial capital considered here. This is not surprising, since a capital growth rate $r$ with values close to zero means that the household will almost surely fall into the area of poverty. On the other hand, these figures also show that when the capital growth rate $r$ reaches higher values (in this case, when $b$ increases), the trapping probability for insured households (with or without subsidies) declines more rapidly to zero compared to that of uninsured households. Similarly, the cost of social protection is also a decreasing function of the rate of income generation $b$, with the only exception that under the scheme with subsidised constant premiums, the cost of social protection converges to $\frac{\beta}{\delta}$, as higher capital growth rates $r$ will almost surely guarantee that households will never fall into the poverty area and therefore governments will be on the need to subsidise premiums indefinitely. Similar results are obtained for the rate of consumption $(0<a<1)$ and the rate on investment or savings $(0<c<1)$, which also have a direct impact on the capital growth rate $r$.

The expected capital loss size has a direct impact on both the expected size of the losses experienced by households and the capital growth rate (for insured households, with or without subsidies, as they pay premiums in exchange of insurance coverage). Figures \ref{Appendix B: Effects of Underlying Factors on the Trapping Probability and the Cost of Social Protection-Figure1}, \ref{Appendix B: Effects of Underlying Factors on the Trapping Probability and the Cost of Social Protection-Figure2} and \ref{Appendix B: Effects of Underlying Factors on the Trapping Probability and the Cost of Social Protection-Figure3} show that when households are expected to experience higher capital losses and will therefore be on the need to pay higher premiums (this is, when $\alpha$ has lower values, but still meets the requirement $b > (1+\theta) \cdot(1-\kappa) \cdot \lambda \cdot \frac{1}{\alpha}$), the trapping probabilities for insured and insured alternatively subsidised households attain high values for the three groups of households considered. Subsidised insured households exhibit a similar behaviour, with the constraint $\pi^{*}\leq (1+\theta) \cdot(1-\kappa) \cdot \lambda \cdot \frac{1}{\alpha}$ providing the admissible range of values for the parameter $\alpha$. The cost of social protection is also a decreasing function of $\alpha$ due to the fact that higher values of $\alpha$ reduce the likelihood of governments requiring to inject capital to lift households out of poverty. Equivalent results are obtained when analysing all parameters that have a direct influence on the dynamics of the capital losses and the capital growth rate, e.g. the expected capital loss frequency $(\lambda>0)$, the proportionality factor $\kappa \in [0,1]$ and the loading factor $(\theta \geq 0)$.

As discussed previously, an initial capital that lies close to the critical capital $x^{*}$ will lead to higher trapping probabilities. In addition, the cost of social protection is an increasing function of the critical capital $x^{*}$. However, the sensitivity analysis shows that this is only true for the uninsured, insured and insured alternatively subsidised households.

\newpage

\typeout{get arXiv to do 4 passes: Label(s) may have changed. Rerun}

\end{document}